\documentclass[a4paper,journal,compsoc]{IEEEtran}
\usepackage[utf8]{inputenc}
\usepackage{pdfpages}
\usepackage[noadjust]{cite}
\usepackage{amsmath,amssymb,amsthm}
\newtheorem{thm}{Theorem}
\newtheorem{lem}{Lemma}
\newtheorem{prop}{Proposition}

\theoremstyle{definition}

\newtheorem{problem}{Problem}

\usepackage{xcolor}
\usepackage{hyperref}
\usepackage{tikz,pgfplots}
\usetikzlibrary{shapes,arrows}
\usetikzlibrary{matrix,positioning}
\usetikzlibrary{shapes.multipart}
\usepackage{neuralnetwork}

\newcommand{\ent}[1]{H(#1)}

\newcommand{\mutinf}[1]{I(#1)}

\newcommand{\kld}[2]{D(#1\Vert#2)}

\newcommand{\naturals}{\mathbb{N}}
\newcommand{\indset}{\mathbb{I}}

\newcommand{\set}[1]{\mathcal{#1}}

\newcommand{\stoch}[1]{\mathbf{#1}}
\newcommand{\Prob}[1]{\mathbb{P}(#1)}
\newcommand{\expec}[1]{\mathbb{E}(#1)}
\newcommand{\mutrate}[1]{\bar{I}(#1)}

\newcommand{\entrate}[1]{\bar{H}(#1)}

\newcommand{\pmf}[1]{p_{#1}}
\newcommand{\tpm}{P}
\newcommand{\agg}{Q}
\newcommand{\invariant}{\pi}
\newcommand{\diag}[1]{\mathrm{diag}(#1)}
\newcommand{\trace}[1]{\mathrm{Tr}(#1)}
\newcommand{\transpose}[1]{#1^{\mathsf{T}}}

\newcommand{\indicator}[1]{\mathbf{1}(#1)}

\title{Information-Theoretic Reduction of Markov Chains}

\author{Bernhard C. Geiger,~\IEEEmembership{Senior Member,~IEEE}%
\IEEEcompsocitemizethanks{\IEEEcompsocthanksitem Bernhard C. Geiger is with Know-Center Gmbh, Graz, Austria. Email: geiger@ieee.org}
}

\IEEEtitleabstractindextext{
\begin{abstract}
We survey information-theoretic approaches to the reduction of Markov chains. Our survey is structured in two parts: The first part considers Markov chain coarse graining, which focuses on projecting the Markov chain to a process on a smaller state space that is \emph{informative} about certain quantities of interest. The second part considers Markov chain model reduction, which focuses on replacing the original Markov model by a simplified one that yields \emph{similar} behavior as the original Markov model. We discuss the practical relevance of both approaches in the field of knowledge discovery and data mining by formulating problems of unsupervised machine learning as reduction problems of Markov chains. Finally, we briefly discuss the concept of lumpability, the phenomenon when a coarse graining yields a reduced Markov model.
\end{abstract}
\begin{IEEEkeywords}
Markov chains, coarse graining, model reduction, lumpability, clustering, community detection, knowledge discovery
\end{IEEEkeywords}
}

\begin{document}

\maketitle

\section{Introduction}
Markov chains are ubiquitous in many scientific disciplines: From $n$-gram models in natural language processing, over computational models for chemical reactions, to models for the behavior of communication channels (including speech communication), the Markov property has enabled analytically tractable and computationally efficient simulation and inference. In certain cases, however, the resulting Markov chains have an extremely large state space, thus preventing us to fully utilize these attractive properties. Examples for scientific fields with large Markov chains are:
\begin{itemize}
 \item Natural language processing, where higher-order Markov chains ($n$-grams) act as language models that quantify the probability that a token (e.g., a word) follows a given sequence of tokens~\cite[Ch.~6]{Manning_NLP}.
 \item Agent-based modeling, where the set of possible agent configurations comprises the state space of the Markov chain, and state transitions correspond to changes in agent configurations~\cite[Ch.~3]{Banisch_ABM}.
 \item Computational chemistry and computational biology, where reactions are modeled as transitions of a (continuous-time) Markov chain, whose state space is determined by the number of possible species configurations~\cite[Ch.~1]{Koeppl_Circuits}.
\end{itemize}
Thus, in these and other fields, there is the need to \emph{simplify} or \emph{reduce} these Markov models. 

The two main approaches to reducing a Markov model considered in this survey are coarse graining and model reduction. In coarse graining, one aims at finding an (small-alphabet) observation of the original Markov chain that remains informative about some quantity of interest (Section~\ref{sec:coarsegraining}). Model reduction, on the other hand, is concerned with replacing a Markov model by a simpler one, with the aim that the simplified or reduced model exhibits a similar behavior as the original one (Section~\ref{sec:modelreduction}). Whereas coarse-graining is focused on realizations of Markov chains and, thus, data compression, model reduction is concerned with the models themselves and is thus related with model compression.

It may appear remarkable that data and model compression are separate problems in Markov chains. Indeed, one may argue that a coarse graining of a Markov chain already implies the existence of a reduced model. This is true only to a limited extent, as an observation of a Markov chain usually does not exhibit the Markov property. Thus, while the number of different observations can be significantly reduced via coarse graining, the resulting stochastic process has longer-ranging temporal dependencies, implying a trade-off between model complexity and data complexity. There is a rare situation, however, where the coarse-grained process possesses the Markov property, and where thus model and data compression can be achieved simultaneously. This rare situation is referred to as \emph{lumpability}, and we review it (together with the principal ideas of coarse graining and model reduction) in Section~\ref{sec:overview}.

\begin{figure*}[t]
 \centering
 \includegraphics[width=0.8\textwidth]{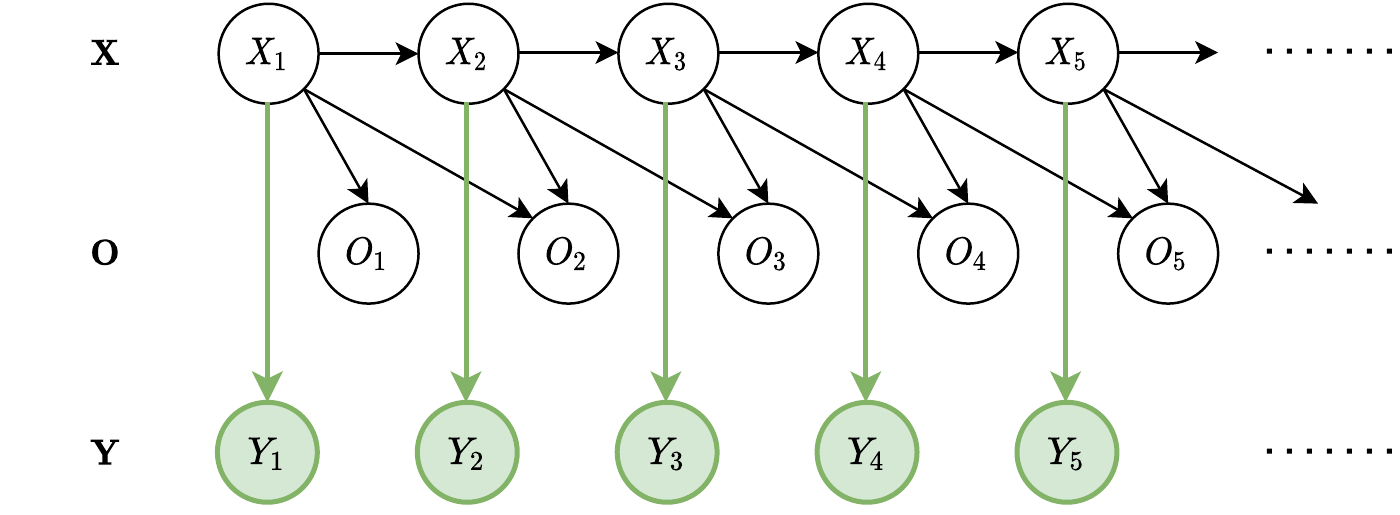}
 \caption{Markov chain coarse graining. Given a Markov chain $\stoch{X}$ and a stochastically related process of interest $\stoch{O}$, the aim is to find a coarse-grained process $\stoch{Y}$ that contains as much information as possible about $\stoch{O}$, cf.~Problem~\ref{prob:coarsegraining}. In the figure, the process $\stoch{O}$ is related to the Markov chain $\stoch{X}$ via the indicated arrows, i.e., $\pmf{\stoch{O}|\stoch{X}}=\prod_{t\in\naturals}\pmf{O_{t+1}|X_{t+1},X_{t}}$. Further, we assume a symbol-wise coarse graining (see Section~\ref{sec:coarsegraining:simple}), where $Y_t$ is conditionally independent from $\stoch{X}$ given $X_t$.}
 \label{fig:coarsegraining}
\end{figure*}

We will show that coarse graining and model reduction can be formulated as optimization problems (cf. Problem~\ref{prob:coarsegraining} and Problem~\ref{prob:modelreduction}, respectively). Solving these optimization problems involves finding patterns within the probabilistic description or within realizations of the Markov chain, which can be used to reduce the model, effectively lowering the computational complexity of operations related to it. But as much as we are interested in the reduced model or coarse-grained process realizations, also the \emph{patterns} enabling this reduction may be relevant. Indeed, these patterns in the probabilistic description or the realization help us get a better understanding of the Markov chain or the real-world process it models. Thus, several of the methods we survey in this work can not only be used for Markov reduction, but also for exploratory analysis of the Markov chain under consideration. Further, since many machine learning problems can be formulated as reduction problems of Markov chains, several of the discussed approaches can be employed to solve (mainly unsupervised) machine learning problems. This suggests that Markov reduction is a viable approach to knowledge discovery and data mining. We will expand on this in Section~\ref{sec:discussion}.

\textbf{Scope.} This work focuses on discrete-time, finite-state Markov chains. We assume that these Markov chains are irreducible, aperiodic, and stationary (see~\cite{Kemeny_FMC} for terminology). This setting, although simple at the first glance, exhibits a rich portfolio of non-trivial properties. Aside from occasional remarks, this survey does not cover reduction of continuous-time Markov chains~\cite{Katsoulakis_CoarseGraining,Petrov_Reaction}, in-homogeneous or non-stationary Markov chains~\cite{Bujorianu_Inhomogeneous}, general Markov processes~\cite{Rosenblatt_MarkovianFunctions}, or hidden Markov models~\cite{White_HMM,Kotsalis_HMMReduction,Deng_HMM}. Further, we focus on approaches to Markov chain coarse graining and Markov model reduction that are information-theoretic, i.e., where the objective function utilizes information-theoretic quantities such as entropy, mutual information, or Kullback-Leibler divergence. While we occasionally provide pointers to literature on other objectives, we do not aim to cover this related literature exhaustively.

\textbf{Notation.} 
All RVs are defined on the common probability space $(\Omega,\mathcal{F},\mathbb{P})$ and assumed to have finite alphabets. For example, the RV $Z{:}\ \Omega\to\set{Z}$ has probability mass function (PMF) $\pmf{Z}$, where $\pmf{Z}(z)=\Prob{Z^{-1}(z)}=\Prob{Z=z}$ holds for every realization $z$ from its alphabet $\set{Z}$. Joint and conditional PMFs of multiple RVs are defined accordingly.

A stochastic process is a sequence of RVs and shall be denoted by a boldface upper case letter, i.e., $\stoch{Z}=(Z_t,\ t\in\naturals)$. We let $\pmf{\stoch{Z}}$ denote the joint PMF of the process. For a sub-sequence of the process indexed by $\indset\subseteq\naturals$ we write $Z_\indset=(Z_t,\ t\in\indset)$ and $\pmf{Z_\indset}$ for its PMF. If $\indset$ is a set of consecutive integers, e.g., $\indset=\{t,t+1,\dots,t+T\}$, then we abuse notation and write $Z_\indset=Z_t^{t+T}$. We further abbreviate $[N]=\{1,\dots,N\}$ for some integer $N\in\naturals$. In this work, we assume that all stochastic processes are stationary, i.e., for every $\indset\subset\naturals$ and every $t\in\naturals$, we have $\pmf{Z_\indset}=\pmf{Z_{\indset+t}}$.

The main focus of this work are Markov chains and observations of Markov chains. For Markov chains, the process distribution factorizes as
\begin{equation}
 \pmf{\stoch{Z}}(\stoch{z}) = \pmf{Z_1}(z_1)\prod_{t\in\naturals} \pmf{Z_{t+1}|Z_{t}}(z_{t+1}|z_{t}).
\end{equation}
If the $\stoch{Z}$ is time-homogeneous, then the alphabet of $X_t$ is the same for all $t$ -- we then call the common alphabet $\set{Z}$ the state space of $\stoch{Z}$. Further, $\pmf{Z_{t+1}|Z_{t}}$ does not depend on $t$ and there exists a transition probability matrix $\tpm$ with entries $\tpm_{z,z'}=\pmf{Z_{t+1}|Z_{t}}(z'|z)$. Unless otherwise noted, we assume that our Markov chains are irreducible and aperiodic (see~\cite{Kemeny_FMC} for terminology), i.e., there exists a unique invariant distribution vector $\pi$ that satisfies $\pi^T=\pi^T\tpm$.

We assume basic understanding of information-theoretic quantities, such as the entropy $\ent{Z}$ of a RV $Z$, the mutual information $\mutinf{Z;Z'}$ between two RVs $Z$ and $Z'$, or the entropy rate $\entrate{\stoch{Z}}$ of a stochastic process $\stoch{Z}$; we refer the reader to~\cite{Cover_Information} for definitions.

The indicator function is denoted as $\indicator{A}$ and is equal to 1 if $A$ is true and to 0 if $A$ is false.

\section{Coarse Graining, Model Reduction, and Lumpability}\label{sec:overview}
In this section, we discuss the problems of Markov chain coarse graining, Markov model reduction, and lumpability, a phenomenon that connects the former two problems. 

\subsection{Coarse Graining}
For the problem of coarse graining, suppose that some Markov chain is informative about a quantity of interest. We are now interested in taking observations of this Markov chain that remain informative about this quantity of interest (see Figure~\ref{fig:coarsegraining}).

\begin{figure*}[t]
 \centering
 \includegraphics[width=.6\textwidth]{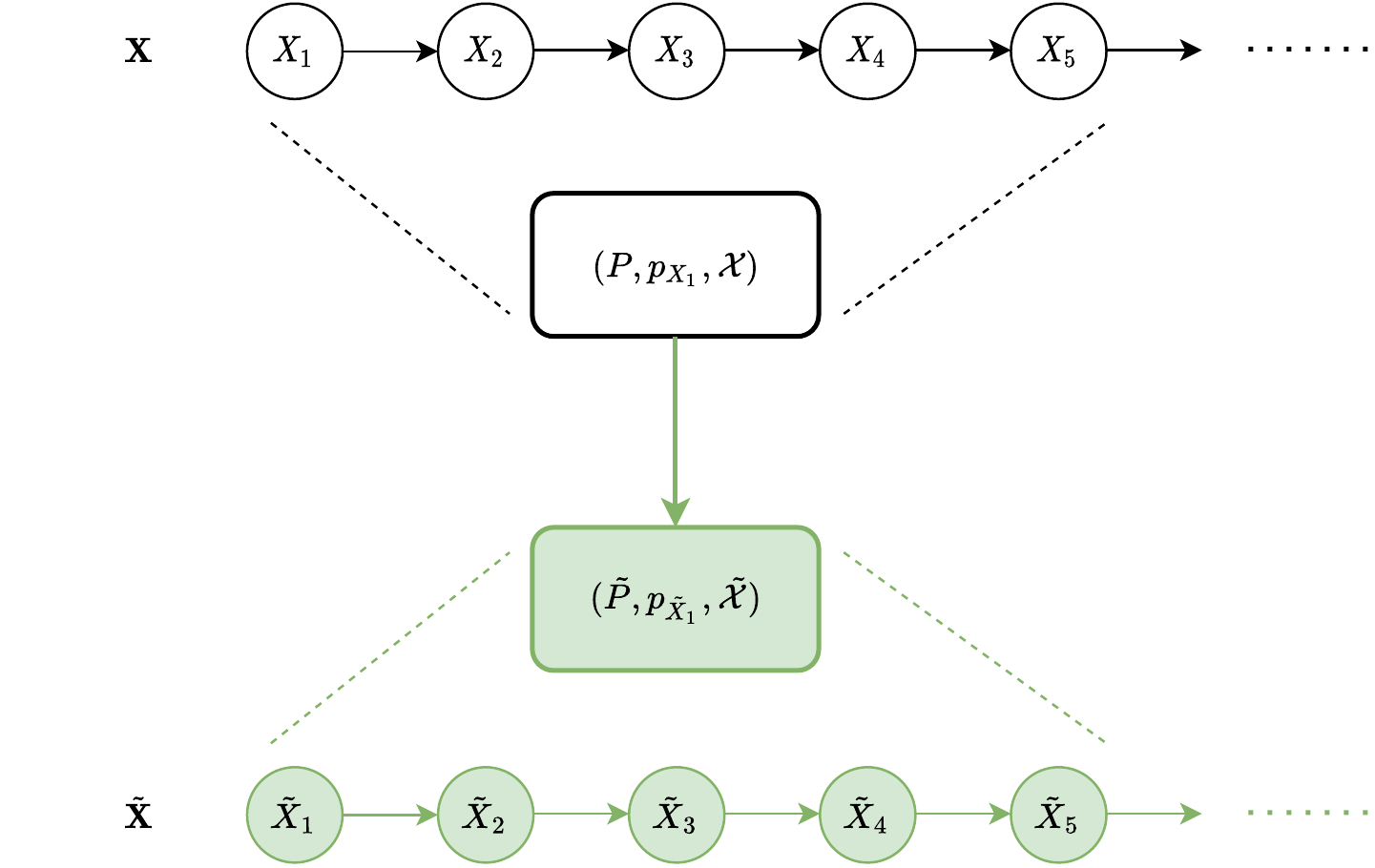}
 \caption{Markov chain model reduction. Given a Markov chain $\stoch{X}$, the aim is to find a reduced model $\tilde{\stoch{X}}$ that is similar to the original model in a well-defined sense, cf.~Problem~\ref{prob:modelreduction}. Here, we consider all types of reductions: Reductions to a smaller state space (i.e., $|\set{X}|\gg|\tilde{\set{X}}|$, cf. Section~\ref{sec:modelreduction:aggregation}), replacing the original transition probability matrix $\tpm$ by one with reduced complexity (Section~\ref{sec:modelreduction:parameters}), or increasing the compressibility of the reduced model (Section~\ref{sec:modelreduction:compressibility}).}
 \label{fig:modelreduction}
\end{figure*}

\begin{problem}[Coarse Graining]\label{prob:coarsegraining}
 Let $\stoch{X}$ be a stationary Markov chain and let the stochastic process $\stoch{O}$, which is jointly stationary with $\stoch{X}$, represent a quantity of interest. The problem of coarse graining is to find a maximizer of the following optimization problem:
 \begin{equation}\label{eq:coarsegraining}
  \max_{\pmf{\stoch{Y}|\stoch{X}}\in\set{P}} \mathsf{Inf}(\stoch{Y}\to \stoch{O})
 \end{equation}
 where $\pmf{\stoch{Y}|\stoch{X}}$ is a conditional distribution determining the observation $\stoch{Y}$ of the Markov chain, where $\set{P}$ is the feasible set, and where $\mathsf{Inf}(\stoch{Y}\to \stoch{O})$ measures the information $\stoch{Y}$ carries about $\stoch{O}$.
\end{problem}

Since, by this assumption, the processes $\stoch{O}$ and $\stoch{Y}$ are conditionally independent given $\stoch{X}$, for any measure of informativeness that satisfies a data processing inequality~\cite[Th.~2.8.1]{Cover_Information},~\eqref{eq:coarsegraining} will be maximized for the identity mapping, i.e., for $\stoch{Y}\equiv\stoch{X}$. Thus, in all non-trivial cases the feasible set $\set{P}$ needs to be restricted. Such restrictions may include limiting the information capacity of the mapping $\pmf{\stoch{Y}|\stoch{X}}$ or the cardinality of the alphabet $\set{Y}$ of $\stoch{Y}$. Other than for the sake of excluding trivial solutions, we may restrict the feasible set for practical purposes. For example, it will often make sense to limit the complexity of the mapping $\pmf{\stoch{Y}|\stoch{X}}$, measured, e.g., by the number of parameters required to describe it. A typical example is the restriction to mappings that factorize, i.e., $\pmf{\stoch{Y}|\stoch{X}}(\stoch{y}|\stoch{x})=\prod_{t\in\naturals} \pmf{Y|X}(y_t|x_t)$, in which case $(\stoch{X},\stoch{Y})$ is a hidden Markov model (HMM,~\cite{Ephraim_HMMs}); if additionally $\pmf{Y|X}(y_t|x_t)=\indicator{y_t=g(x_t)}$ for some function $g{:}\ \set{X}\to\set{Y}$, then $(\stoch{X},\stoch{Y})$ is a functional HMM or a \emph{lumping}. We will discuss information-theoretic approaches to Markov chain coarse graining in Section~\ref{sec:coarsegraining}.

\textbf{Relevance and Sample Applications.}
The concept of coarse graining is strongly related to quantization and clustering, both of which are important in systems design and data exploration, respectively. Coarse graining of Markov chains thus has relevance in fields where the systems under investigation have Markovian inputs, or where the data is (or can be converted to) a random walk. Indeed, information-theoretic objectives for coarse graining have been proposed for random walk-based clustering~\cite{Alush_PairwiseClustering,Tishby_MarkovRelaxation}, semi-supervised clustering~\cite{Steger_SemiSupervised}, co-clustering~\cite{Dhillon_ITClustering,BloechlEtAl_MCClustering} and community detection~\cite{Faccin_CD,Lambiotte_RandomWalks}.

\subsection{Model Reduction}

The problem of model reduction again assumes some Markov chain $\stoch{X}$. The task of model reduction is to obtain a Markov chain $\tilde{\stoch{X}}$ that is stochastically similar to the original one, $\stoch{X}$ (see Figure~\ref{fig:modelreduction}).

\begin{problem}[Model Reduction]\label{prob:modelreduction}
 Let $\stoch{X}$ be a stationary Markov chain. The problem of model reduction is to find a maximizer of the following optimization problem:
 \begin{equation}\label{eq:modelreduction}
  \max_{\tilde{\stoch{X}}\in\set{P}} \mathsf{Sim}(\stoch{X}\to \tilde{\stoch{X}})
 \end{equation}
 where $\tilde{\stoch{X}}$ is a Markov chain, $\set{P}$ is the feasible set, and where $\mathsf{Sim}(\stoch{X}\to \tilde{\stoch{X}})$ measures similarity.
\end{problem}

Any reasonable measure of similarity will be maximized if the entities it compares are identical. Thus, to exclude the trivial case of $\tilde{\stoch{X}}\equiv \stoch{X}$, the feasible set $\set{P}$ will often be restricted. Typical examples for such a restriction may include enforcing a strictly smaller alphabet $\tilde{\set{X}}$ of $\tilde{\stoch{X}}$ than $\set{X}$ or otherwise reducing the number of parameters to describe the stochastic behavior of $\tilde{\stoch{X}}$.

Note further that model reduction for Markov chains automatically includes model reduction for HMMs with finite observation spaces. Indeed, if $\stoch{X'}$ and $\stoch{O}$ are the state and observations process of a HMM, i.e., if $\stoch{X'}$ is Markov and if $\pmf{\stoch{O}|\stoch{X}'}=\prod_{t\in\naturals} \pmf{O_t|X_t'}$, then $\stoch{X}=(\stoch{X}',\stoch{O})$ is a Markov chain. In such a setting, the similarity measure $\mathsf{Sim}$ in~\eqref{eq:modelreduction} can be adjusted to have a stronger focus on approximating the state or the observations process' behavior, respectively. We will discuss information-theoretic approaches to Markov chain model reduction in Section~\ref{sec:modelreduction}.

\begin{figure*}[t]
 \centering
 \includegraphics[width=.9\textwidth]{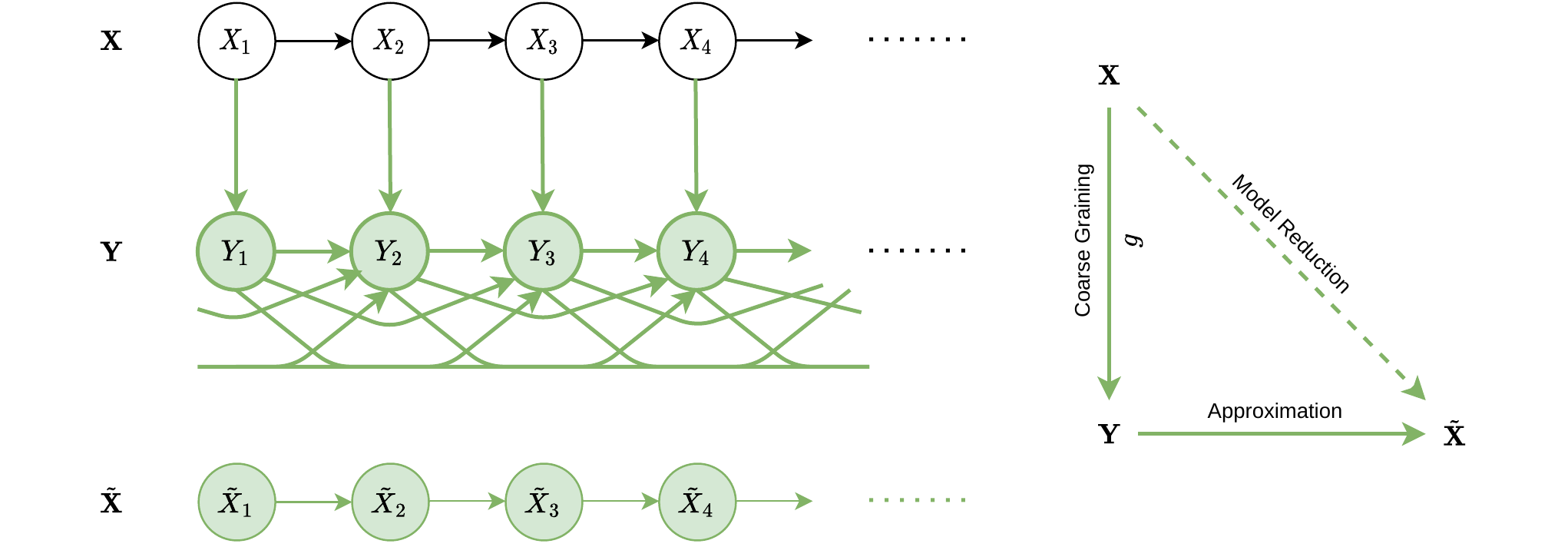}
 \caption{Coarse graining-based model reduction. Given a Markov chain $\stoch{X}$, the aim is to find a reduced model $\tilde{\stoch{X}}$ that is similar to a coarse graining $\stoch{Y}$ of the original model $\stoch{X}$, cf.~Problem~\ref{prob:coarsereduction}. While the coarse graining $\stoch{Y}$ may have more complicated temporal dependence structures (see l.h.s.), coarse graining-based model reduction approximates $\stoch{Y}$ by a Markov model $\tilde{\stoch{X}}$. The r.r.s.\ of the figure is an adaption from~\cite[Fig.~1]{Amjad_GeneralizedMA} and illustrates the interplay between coarse graining and model reduction in this case.}
 \label{fig:interplay}
\end{figure*}

\textbf{Relevance and Sample Applications.}
Model reduction for Markov chains is useful whenever simulating or storing the original Markov model is computationally too complex, or whenever there is too little data to reliably infer the parameters of the original model. For example, reduced models can be useful for running (otherwise costly) agent-based simulations~\cite{Lamarche-Perrin_ABM,Banisch_ABM,Khudabukhsh_LumpabilityABMs}, credit risk modeling~\cite{Georgiou_MarkovCreditRisk}, reducing combinatorial chemical reaction networks~\cite{Petrov_Reaction} (see also~\cite{Tribastone_Simulation}), or for simulating large Markov chains at least approximately. Further, it can be shown that the task of model reduction has the potential to reveal interesting structure in the original Markov chain, a fact that has been utilized, e.g., for community detection~\cite{Rosvall_Infomap,Toth_PAM,Piccardi_CD}.

\subsection{Lumpability: Bridging Coarse Graining and Model Reduction}
\label{sec:overview:lumpability}
As we have outlined in the introduction, coarse graining and model reduction may be problems with conflicting objectives. We now formulate a problem that is a synergy between and that bridges Problems~\ref{prob:coarsegraining} and~\ref{prob:modelreduction}. Specifically, we aim for a reduced model that is most similar to a coarse graining of the original Markov chain. 

\begin{problem}\label{prob:coarsereduction}
  Let $\stoch{X}$ be a stationary Markov chain. The problem of lumpability is to find a maximizer of the following optimization problem:
 \begin{equation}\label{eq:coarsereduction}
  \max_{(\tilde{\stoch{X}}, \pmf{\stoch{Y}|\stoch{X}}) \in\set{P}} \mathsf{Sim}(\stoch{Y} \to \tilde{\stoch{X}})
 \end{equation}
 where $\tilde{\stoch{X}}$ is a Markov chain, $\pmf{\stoch{Y}|\stoch{X}}$ is a conditional distribution determining the observation $\stoch{Y}$ of the Markov chain, $\set{P}$ is the feasible set, and where $\mathsf{Sim}(\stoch{Y} \to \tilde{\stoch{X}})$ measures the similarity of the processes  $\stoch{Y}$ and $\tilde{\stoch{X}}$.
\end{problem}

As before, we have to install restrictions to prevent the problem from becoming trivial (as in cases in which $\pmf{\stoch{Y}|\stoch{X}}$ is the identity mapping and $\tilde{\stoch{X}}\equiv\stoch{Y}\equiv\stoch{X}$). Again, typical restrictions apply to the cardinality of the alphabet $\set{Y}$ of $\pmf{\stoch{Y}|\stoch{X}}$, while the optimization over the Markov chain $\tilde{\stoch{X}}$ can remain unconstrained.

We have stated this problem in large generality. Let us restrict the coarse graining $\pmf{\stoch{Y}|\stoch{X}}$ to be symbol-by-symbol (i.e., $\pmf{\stoch{Y}|\stoch{X}}$ factorizes as $\pmf{Y|X}$) and deterministic (i.e., $\pmf{Y|X}$ is defined by a coarse graining function $g$). With this, we enter the domain of \emph{lumpability}~\cite[\S6.3]{Kemeny_FMC}. Indeed, a Markov chain $\stoch{X}$ is \emph{lumpable} w.r.t.\ a coarse graining function $g$ or w.r.t.\ its induced partition if the coarse-grained process $\stoch{Y}$ is a Markov chain. This situation is quite rare, as for a given function $g$, the set of \emph{lumpable} Markov chains is a null set~\cite[Th.~31]{GurvitsLedoux_MarkovPropertyLinearAlgebraApproach}. 

Nevertheless, lumpability is an attractive property from the perspective of computation. For example, if $\pi$ is the invariant distribution of a Markov chain $\stoch{X}$ lumpable w.r.t.\ $g$, then the invariant distribution of the reduced Markov chain $\tilde{\stoch{X}}$ coincides with the coarse graining of $\pi$ by $g$~\cite[Th.~4]{Buchholz_Exact}; similar results can be shown to connect the marginal distributions $\pmf{X_t}$ and $\pmf{\tilde X_t}$. 

Lumpability\footnote{There are multiple definitions of lumpability; most notably, strong or ordinary lumpability~\cite[\S6.3]{Kemeny_FMC}, weak lumpability~\cite[\S6.4]{Kemeny_FMC}, and exact lumpability~\cite{Buchholz_Exact}. We will focus on strong lumpability in this work.} has been studied extensively in the past: Investigations have been made for general Markov processes~\cite{Rosenblatt_MarkovianFunctions}, continuous-time Markov chains~\cite{Tian_CTLumpability}, discrete-time Markov chains~\cite{Burke_Lumpability}, for lumpability of higher order (both from information-theoretic~\cite{GeigerTemmel_kLump} and linear-algebraic perspectives~\cite{GurvitsLedoux_MarkovPropertyLinearAlgebraApproach}), and for Markov random fields~\cite{Geiger_MRFs}. Further, it has been investigated when the state process of a hidden Markov model can be coarse grained without losing the hidden Markov property~\cite{White_HMM}, and when a stochastic coarse graining $\pmf{Y|X}$ of a Markov chain has the Markov property~\cite{Spreij_HMM}.

Let us consider again Problem~\ref{prob:coarsereduction}. For deterministic coarse grainings $g$, the problem can be rewritten as a double maximization problem, with the aim of finding maximizers of
 \begin{equation}\label{eq:coarsereduction_2}
  \max_{g \in\set{P}'} \max_{\tilde{\stoch{X}} \in\set{P}''} \mathsf{Sim}(\stoch{Y} \to \tilde{\stoch{X}})
 \end{equation}
If the chosen coarse graining $g$ is such that $\stoch{X}$ is lumpable w.r.t.\ it, then $\stoch{Y}$ is a Markov chain, and the closest Markov model $\tilde{\stoch{X}}$ to it is simply the stochastic description of $\stoch{Y}$. The inner maximization over $\tilde{\stoch{X}}$ in~\eqref{eq:coarsereduction_2} thus becomes trivial. If the Markov chain $\stoch{X}$ is not lumpable w.r.t.\ to any function $g$, then the maximization over $\tilde{\stoch{X}}$ depends on the selected similarity measure. In such cases it may still be relevant to search for \emph{quasi- or approximately lumpable} coarse grainings of a Markov chain. Fig.~\ref{fig:interplay} illustrates the interplay between coarse graining and model reduction that is motivated by the phenomenon of lumpability. We will present information-theoretic approaches towards finding (approximately) lumpable coarse grainings in Section~\ref{sec:coarsegraining:simple:lumpable}. The approximation of the resulting coarse-grained process by a Markov chain is covered in Section~\ref{sec:modelreduction:aggregation}.

\section{Information-Theoretic Approaches to Markov Chain Coarse Graining}\label{sec:coarsegraining}
We now survey the literature on information-theoretic coarse graining. In Section~\ref{sec:coarsegraining:rd} we start our analysis in a setting without an observation process $\stoch{O}$ and with an unrestricted coarse graining map $\pmf{\stoch{Y}|\stoch{X}}$, and we connect the resulting problem to rate-distortion theory. We then restrict the coarse graining to be symbol-by-symbol in Section~\ref{sec:coarsegraining:simple}, a simple, but very rich setting. We finally introduce the observation process $\stoch{O}$ in Section~\ref{sec:coarsegraining:observation}.

\subsection{Block-Wise Coarse Graining and Rate-Distortion Theory}\label{sec:coarsegraining:rd}

Measuring informativeness by the distortion between $\stoch{X}$ and a reconstruction $\hat{\stoch{X}}$ obtained from the coarse graining $\stoch{Y}$, and restricting the feasible set by limiting the information that $\stoch{Y}$ shares with $\stoch{X}$ leads us to the field of rate-distortion theory~\cite[Ch.~13]{Cover_Information}. Specifically, let $\hat{\stoch{X}}$ assume values in $\set{X}$ and let $d_H{:}\ \set{X}^2\to\{0,1\}$ denote the Hamming distortion, i.e., $d_H(x,\hat x)=\indicator{x\neq \hat x}$. For vectors $x_1^T$ and $\hat{x}_1^T$ we abuse notation and write $d_H(x_1^T,\hat{x}_1^T)=\frac{1}{T}\sum_{\ell=1}^T d_H(x_\ell,\hat{x}_\ell)$. Then, rate-distortion theory is concerned with finding a sequence of minimizers of
\begin{equation}\label{eq:rate-distortion}
 \lim_{T\to\infty} \min_{\pmf{\hat{X}_1^T|X_1^T}} \frac{1}{T} \mutinf{X_1^T;\hat{X}_1^T}
\end{equation}
where the minimizations are performed over sets of conditional distributions satisfying $\expec{d_H(X_1^T,\hat{X}_1^T)}\le D$ for some distortion constraint $D$. The minimal value that~\eqref{eq:rate-distortion} assumes for a given distortion $D$ is called the rate-distortion function at $D$ and abbreviated $R(D)$. The problem greatly simplifies if $\stoch{X}$ is iid, in which case the minimizing mappings $\pmf{\hat X|X}$ can be found via the Blahut-Arimoto algorithm~\cite[Ch. 13.8]{Cover_Information}. Outside of the iid case, however, the problem is largely unsolved. For $\stoch{X}$ being a Markov chain with binary state space and symmetric state transition probabilities, bounds on the rate-distortion function have been presented~\cite{Jalali_SymmetricMarkov}. Further, for small $D$, the corresponding rate-distortion function is known exactly~\cite[eq.~(43)]{Gray_ARProcesses}. 

Note that this informational formulation of the rate-distortion problem does not make the coarse graining $\pmf{\stoch{Y}|\stoch{X}}$ explicit. The corresponding operational formulation, however, does. Namely, the operational formulation seeks sequences of encoder and decoder functions $g_n{:}\ \set{X}^n\to\set{Y}$ and $f_n{:}\ \set{Y}\to\set{X}^n$, respectively, such that $f_n\circ g_n$ leads to an expected distortion less than $D$, cf.~\cite[p.~340]{Cover_Information}. The cardinality of $\set{Y}$ increases with $n$, and the rate of increase is given by the rate-distortion function $R(D)$~\cite[p.~341]{Cover_Information}. In this operational formulation, the mapping $\pmf{\stoch{Y}|\stoch{X}}$ is deterministic, but operates on blocks of length $n$, i.e., we have
\begin{equation}
 \pmf{\stoch{Y}|\stoch{X}} (\stoch{y}|\stoch{x}) = \prod_{\ell} \indicator{y_\ell=g_n(x_{(\ell-1)n+1}^{\ell n})}.
\end{equation}
Note that here $\stoch{Y}$ operates at one $n$-th of the rate as $\stoch{X}$, i.e., the scalar process $\stoch{Y}$ is jointly stationary with the $n$-blocked process $\stoch{X}$ of $n$-dimensional RVs. Note further that in the edge case of a Hamming distortion constraint $D=0$, we are in the lossless compression setting, for which it is known that the entropy rate $\entrate{\stoch{X}}$ assumes the position of the rate-distortion function $R(D)$ as a fundamental limit: Taking $D=0$ in~\eqref{eq:rate-distortion} implies $\hat{X}_1^t=X_1^t$ almost surely, in which case there is no minimization over $\pmf{\hat{X}_1^t|X_1^t}$ and $\mutinf{X_1^t;\hat{X}_1^t}=\ent{X_1^t}$. While the informational formulation thus becomes trivial, the operational formulation is still an instance of Markov chain coarse graining, again with sequences of encoder and decoder functions $g_n$ and $f_n$, operating on blocks of realizations.

\subsection{Symbol-Wise Coarse Graining without Observations}\label{sec:coarsegraining:simple}

Both the lossy and the lossless settings assume long blocklengths $n$ to achieve efficient coding. Long blocklengths, however, require large codebooks for the encoder $g_n$ and decoder $f_n$. We will thus now investigate the special case where the encoder operates symbol-by-symbol, i.e., for a blocklength $n=1$. We further ignore the reconstruction problem, i.e., from the Markov tuple $\stoch{X} - \stoch{Y} - \hat{\stoch{X}}$ we only consider the first part. Restricting the optimization to functions $g{:}\ \set{X}\to \set{Y}$ and writing $\stoch{Y}=(g(X_t),\ t\in\naturals)$, we thus consider the following minimization problem:
\begin{equation}\label{eq:coarsegraining:symbol}
 \min_{g} \mathsf{Inf}(\stoch{Y}\to \stoch{X})
\end{equation}
In the remainder of this section, we will investigate several ways to measure information in~\eqref{eq:coarsegraining:symbol}.

\subsubsection{Preserving Process Information}

Measuring the informativeness by the information loss rate $\entrate{\stoch{X}|\stoch{Y}}$ was investigated in~\cite{Watanabe_Infoloss}, where it was shown that $\entrate{\stoch{X}|\stoch{Y}}\le\ent{X|Y}$~\cite[Th.~3]{Watanabe_Infoloss} and that a nontrivial coarse graining of a process can satisfy $\entrate{\stoch{X}|\stoch{Y}}=0$ due to the redundancy inherent in $\stoch{X}$, cf.~\cite[Sec.~V]{Watanabe_Infoloss}. Geiger and Temmel investigated this case for Markov chains, i.e., they characterized the szenario in which there exists a $g$ such that $\entrate{\stoch{X}|\stoch{Y}}=0$. To this end, let $y_1^T$ be a coarse-graining of a realization $x_1^T$ obtained via a $g{:}\ \set{X}\to\set{Y}$, and let $\set{R}(y_1^T)$ denote the set of realizable preimages, i.e., $\set{R}(y_1^T)=\{x_1^T\in\set{X}^T{:}\ \pmf{X_1^T}(x_1^T)>0,\ \forall t\in[T]{:}\ y_t=g(x_t)\}$. Then, $\entrate{\stoch{X}|\stoch{Y}}=0$ if and only if $|\set{R}(Y_1^T)|$ is almost surely uniformly bounded for all $T$~\cite[Th.~1]{GeigerTemmel_kLump}.\footnote{A similar analysis for automata was conducted in~\cite{Crespi_Automata}.}. From this immediately follows that the \emph{absolute} information loss caused by the coarse graining is finite (and not just sub-linear) in $T$.
\begin{prop}[{\cite[Prop.~1]{GeigerTemmel_kLump}}]\label{prop:klump:lossbound}
 If $\entrate{\stoch{X}|\stoch{Y}}=0$, then for every $T$ we have $\ent{X_1^T|Y_1^T}\le 2\log(|\set{X}|-|\set{Y}|+1)$.
\end{prop}
Despite the fact that $\stoch{Y}$ is quite informative about $\stoch{X}$ in the sense that $\entrate{\stoch{Y}}=\entrate{\stoch{X}}$, perfect reconstruction may not be possible without some (finite amount of) additional side information. In other words, while the encoder function $g$ is informative, it may be difficult or even impossible to obtain a decoder map $f_T$ ensuring that $d_H(X_1^T, f_T(g(X_1^T)))$ is small.

This characterization of information preservation in~\cite[Th.~1]{GeigerTemmel_kLump} is \emph{structural}, as it depends only on whether an entry in the transition probability matrix $\tpm$ is positive or not (i.e., it depends on the the transition graph of the Markov chain), but not on the actual value. For example, based on the properties of the transition graph, a bound on the alphabet size $|\set{Y}|$ can be derived above which information-preserving coarse grainings are possible~\cite[Cor.~1]{Geiger_Markov_arXiv}. In essence, information-preserving coarse graining is only possible in a sufficiently sparse setting; if $\tpm$ is positive, then $\entrate{\stoch{X}|\stoch{Y}}>0$ for any nontrivial coarse graining~\cite[Cor.~1]{GeigerTemmel_kLump}.

Generally, even for a given transition graph it is hard to check whether a given function $g$ yields an information-preserving coarse graining. More specifically, the complexity of deciding whether $\entrate{\stoch{X}|\stoch{Y}}=0$ is exponential in $|\set{X}|^2$, cf.~\cite[Sec.~2.6]{GeigerTemmel_kLump}. These complications can be partly alleviated by certain sufficient conditions for a function $g$ to provide an information-preserving coarse graining for a given Markov chain $\stoch{X}$. One of these sufficient conditions is the \emph{single entry} property~\cite[Def.~3]{GeigerTemmel_kLump}, which can be checked with a computational complexity of $\mathcal{O}(|\set{X}|^2)$. Furthermore, for a given Markov chain $\stoch{X}$, a function $g$ satisfying the single entry property can be found via solving a clique partition problem~\cite{GeigerHoferTemmel_ZEGraph_arXiv}. The benefit of this sufficient condition comes at the cost of a more stringent requirement on the alphabet size of the coarse graining or, equivalently, on the sparsity of $\tpm$. More specifically, the tuple $(\tpm,g)$ can only be single entry if~\cite[Prop.~2]{GeigerHoferTemmel_ZEGraph_arXiv} 
\begin{equation}
 |\set{Y}|\ge \max_{x\in\set{X}} \sum_{x'\in\set{X}} \indicator{\tpm_{x,x'}>1}.
\end{equation}
 If the transition graph is not sparse but if sufficiently many of its edges have low probability, then the information loss caused by coarse graining can be bounded, cf.~\cite[Prop.~3]{GeigerHoferTemmel_ZEGraph_arXiv}.

Finally, suppose that a coarse graining $g$ induces an information loss of $\entrate{\stoch{X}|\stoch{Y}}$, and suppose that $\gamma$ is a refinement of $g$, i.e., every preimage under $\gamma$ is contained within a preimage under $g$. Then, the information loss induced by $\gamma$ satisfies
\begin{align*}
 \entrate{\stoch{X}|\gamma(\stoch{X})} &= \lim_{T\to\infty}\frac{1}{T} \ent{X_1^T|\gamma(X_1^T)} \\
 &= \lim_{T\to\infty}\frac{1}{T} \ent{X_1^T|\gamma(X_1^T),Y_1^T} \\
 &\le \lim_{T\to\infty}\frac{1}{T} \ent{X_1^T|Y_1^T} = \entrate{\stoch{X}|\stoch{Y}}
\end{align*}
where the second equality follows from the fact that $g(x)$ is a function of $\gamma(x)$ for every $x\in\set{X}$ and where the inequality follows from the fact that conditioning reduces entropy~\cite[Th.~2.6.5]{Cover_Information}. From this immediately follows that if $g$ is an information-preserving coarse graining, then so are all its refinements. That refinements of single entry coarse grainings remain single entry follows from the fact that a clique can be partitioned in cliques.

\subsubsection{Preserving the Temporal Dependence Structure}

Another type of informativeness in the sense of Problem~\ref{prob:coarsegraining} considers the temporal structure of the original Markov chain $\stoch{X}$. Here, rather than requiring that the coarse-grained process $\stoch{Y}$ allows reconstructing the original chain $\stoch{X}$ with low distortion, the aim is to ensure that the coarse-grained process $\stoch{Y}$ retains the temporal dependence of $\stoch{X}$ --- if $\stoch{X}$ is predictable to some extent, then so should be $\stoch{Y}$.

More concretely, let $\lim_{t\to\infty}\mutinf{X_t;X_1^{t-1}}=\ent{X}-\entrate{\stoch{X}}$ denote the redundancy rate of a stationary process. It can be shown, e.g.,~\cite[Th.~5]{Verdu_GeneralizingFano}, that this redundancy rate provides bounds the error of estimating the next state of the process $X_t$ from its entire past. Thus, preserving the temporal dependence structure of the coarse-grained process amounts to finding a function $g$ that maximizes the redundancy rate $\ent{Y}-\entrate{\stoch{Y}}$. For a Markov chain $\stoch{X}$, this redundancy rate computes to $\mutinf{X_1;X_2}$. Since a coarse graining of a Markov chain usually does not retain the Markov property, we potentially have to maximize a quantity that involves a limit; the entropy rate of a function of a Markov chain is notoriously difficult to compute~\cite{Blackwell_HMMRate}. However, as the data processing inequality or the monotonicity property of mutual information suggests, we can bound the redundancy rate of $\stoch{Y}$ from below by the quantity $\mutinf{Y_1;Y_2}$.

Maximizing this quantity was proposed by~\cite{Meyn_MarkovAggregation,Vidyasagar_MarkovAgg,Deng_ITFramework}, who focused on nearly completely decomposable Markov chains. In such Markov chains, the state space is partitioned into groups within which the Markov chain transitions frequently, but between which transitions only occur rarely. The authors showed that relaxing the combinatorial problem of maximizing $\mutinf{Y_1;Y_2}$ leads to coarse-graining via the Fiedler vector~\cite[Th.~5]{Meyn_MarkovAggregation}, and thus to the spectral theory of Markov chains. A better approximation of the redundancy rate of $\stoch{Y}$ is obtained by computing $\mutinf{Y_k|Y_1^{k-1}}$, for any $k>2$. This approach was taken in~\cite{GeigerWu_HigherOrder} and was shown to yield more appropriate coarse grainings $g$ in a toy example from natural language processing.

The authors of~\cite{Faccin_CD} considered maximizing $\mutinf{Y_t;Y_{t+T}}$ for coarse graining, i.e., they introduced an additional parameter $T$. Their application was community detection in graphs, and they connected this problem to Markov chain coarse graining by letting $\stoch{X}$ be a random walk on said graph. For $T=1$ and for an unweighted, undirected graph, maximizing $\mutinf{Y_t;Y_{t+1}}$ is equivalent to maximizing the likelihood of a degree-corrected stochastic block model; $T>1$ is shown to yield more appropriate results if the graph is sparse and gives rise to long-range path structures. Since $\mutinf{Y_t;Y_{t+T}}$ is trivially maximized by setting $g$ to be the identity function, the authors suggest to either select $g$ from $\set{X}\to[M]$ or to regularize the optimization problem~\cite[eq.~(4)]{Faccin_CD}
\begin{equation}\label{eq:Faccin}
 \max_{g} \mutinf{Y_t;Y_{t+T}} - \beta\ent{Y_t}
\end{equation}
where $\beta\in(0,1)$.

We briefly stick to the setting of community detection and note that a random walk on a graph with a strong community structure is nearly completely decomposable. In other words, for the resulting Markov chain the next community will likely coincide with current community, and the coarse-grained process $\stoch{Y}$ will be predictable if the coarse graining $g$ coincides with the community structure. This setting was investigated in~\cite{Lambiotte_RandomWalks}, where the authors considered a random walk $\stoch{X}$ with transition probability matrix $\tpm$ defined by a connected, undirected graph. With $\invariant$ being the invariant distribution of $\stoch{X}$, the authors defined the clustered autocovariance matrix of $\stoch{X}$ as~\cite[eq.~(3)]{Lambiotte_RandomWalks}
\begin{equation}\label{eq:lambiotte:covariance}
    R_T(V) = \transpose{V} \left(\diag{\invariant}\tpm^T-\transpose{\invariant}\invariant\right)V
\end{equation}
where $V$ is a $|\set{X}|\times M$ coarse graining matrix representing the function $g{:}\ \set{X}\to [M]$ via
\begin{equation}\label{eq:coarsegrainingmatrix}
 V_{\ell,i} =\begin{cases}
              1, & \text{if } g(\ell)=i\\0,&\text{else.}
             \end{cases}
\end{equation}
The authors then aimed at a coarse graining $g$ such that this clustered autocovariance is maximized, i.e., they suggested solving
\begin{equation}\label{eq:lambiotte:opt}
 \max_V \max_{1\le s\le T} \trace{R_s(V)}.
\end{equation}
The coarse-grained process $\stoch{Y}$ should thus have high autocovariance, which is a second-order statistical equivalent to the cost function in~\eqref{eq:Faccin}. The process $\stoch{Y}$ is thus informative about the movement between groups of states (equivalently, about the community structure of the network). Interestingly, for $T=1$,~\eqref{eq:lambiotte:opt} becomes equivalent to modularity maximization, e.g.,~\cite{Newman_Modularity}. For a directed graph and $T=1$, optimizing the corresponding clustered autocovariance becomes equivalent to optimizing the map equation proposed by~\cite{Rosvall_Infomap}, see discussion after~\cite[eq.~(34)]{Lambiotte_RandomWalks}.

\subsubsection{Preserving the Markov Property}
\label{sec:coarsegraining:simple:lumpable}
Several authors addressed the problem of finding a coarse graining function $g$ such that $\stoch{Y}$ is \emph{as Markov as possible} (in other words, the original Markov chain $\stoch{X}$ should be \emph{quasi-lumpable} w.r.t.\ the coarse graining $g$). For example, the authors of~\cite{Derisavi_Lumpability} utilized the original definition of lumpability from~\cite[Th.~6.3.2]{Kemeny_FMC} to derive an algorithm for determining the coarsest coarse graining $g$ (i.e., the smallest $M$) for which $\stoch{X}$ is lumpable. Jacobi~\cite{Jacobi_SpectralLumpability} relies on the fact that if a Markov chain $\stoch{X}$ is lumpable w.r.t.\ a coarse graining $g$ with range $[M]$, that then the transition probability matrix has exactly $M$ eigenvectors that are piecewise constant on the preimage of $[M]$ under $g$. Based on this insight, an appropriate approach to finding a (quasi-)lumpable coarse graining is to simply cluster dimensions of the respective eigenvectors.\footnote{Numerical instabilities in computing eigenvectors can be accounted for by replacing the transition probability matrix by different matrices, such as the invariance matrix proposed in~\cite{Jacobi_SpectralLumpability}.} A completely decomposable Markov chain with $M$ components is lumpable, and its transition probability matrix $\tpm$ has $M$ unit eigenvalues, while the remaining $N-M$ eigenvalues are bounded away from one. Nearly completely decomposable Markov chains can thus be aggregated based on the sign structure (cf.~\cite[Lem.~2.5]{Deuflhard_Identification}) of the $M-1$ non-trivial, approximately piecewise constant right-eigenvectors of $\tpm$ corresponding to the unit eigenvalues, cf.~\cite[pp.~33]{Stewart_MarkovChains}. The authors of~\cite{Zhang_Spectral} also used spectral methods to obtain quasi-lumpable (and quasi-aggregable) coarse grainings. More specifically, they proposed to approximate the matrix of the $M$ leading left or right singular vectors of the empirical transition probability matrix $\tpm$ by a matrix with columns that are piecewise constant on the candidate partition. This approximation is done w.r.t.\ the squared Frobenius norm, which renders the combinatorial problem of simultaneously optimizing the candidate partition and the approximating matrix similar to k-means with squared Euclidean distance. The authors provide bounds on the classification error for certain classes of Markov chains~\cite[Th.~6 \& 7]{Zhang_Spectral}. 

The authors of~\cite{GeigerEtAl_OptimalMarkovAggregation} relied on the information-theoretic characterization of strong lumpability from~\cite[Th.~2]{GeigerTemmel_kLump}, stating that the coarse-grained process $\stoch{Y}$ is a $k$-th order Markov chain if $\ent{Y_k|Y_2^{k-1},X_0}=\ent{Y_k|Y_0^{k-1}}$.\footnote{The concept of strong lumpability, including the information-theoretic characterization for $k=1$, was called \emph{information closure} in~\cite{Pfante_LevelID}.} Seeking a coarse graining $g$ yielding a first-order Markov chain, Geiger et al.\ thus proposed to find a minimizer of
\begin{equation}\label{eq:lumpability_optimization}
 \min_{g\in \set{P}_M} \ent{Y_2|Y_1}-\ent{Y_2|X_1}
\end{equation}
where $\set{P}_M$ is the set of all conditional distributions for a RV with alphabet size $M$.\footnote{The set $\set{P}_M$ obviously includes all functions from $\set{X}$ to $[M]$. We will thus simplify notation and write $g\in\set{P}_M$ to indicate that the optimization is constrained over functions with range $[M]$.}
A relaxation of~\eqref{eq:lumpability_optimization} is an instance of the information bottleneck (IB) problem~\cite{Tishby_InformationBottleneck}, and the authors of~\cite{GeigerEtAl_OptimalMarkovAggregation} proposed using the agglomerative approach of~\cite{Slonim_Agglomerative} to coarse grain the Markov chain $\stoch{X}$. Coarse grainings to higher-order Markov chains were considered in~\cite{GeigerWu_HigherOrder}.

Geiger and Temmel discovered that a subset of \emph{single entry} coarse grainings also yields processes $\stoch{Y}$ that are Markov chains of higher order, cf.~\cite[Def.~4 \& Prop.~5]{GeigerTemmel_kLump}. In other words, there are structural properties depending only on whether transition probabilities are zero or positive, guaranteeing that the coarse-grained process $\stoch{Y}$ is information preserving and a Markov chain of a given order. An algorithm obtaining all such coarse grainings for a given Markov chain $\stoch{X}$ is presented in~\cite{Geiger_Markov_arXiv}.

\subsubsection{Other Measures of Informativeness}
A popular approach to community detection, the Infomap method~\cite{Rosvall_Infomap}, can also be formulated as a coarse graining problem. In this formulation, one aims to find a coarse graining $g$ such that the original Markov chain $\stoch{X}$ is easily compressible w.r.t.\ this coarse graining. Since the authors aim for lossless compression, the entropy rate $\entrate{\stoch{X}}$ is a fundamental limit for the average number of bits required to represent a single realized symbol, cf.~Section~\ref{sec:coarsegraining:rd}. If we want to come close to this limit via symbol-by-symbol encoding, we require a complicated codebook structure, i.e., each state in the Markov chain requires its own codebook, having a separate codeword for every outgoing transition. At the other extreme, a single codebook for symbol-by-symbol encoding requires at least $\ent{X}$ bits per symbol (on average)~\cite[Ch.~5]{Cover_Information}. The Infomap method proposes a codebook structure for symbol-by-symbol encoding that lies between those two extremes. Specifically, assuming a candidate coarse graining $g$, Infomap uses separate codebooks for each group of states and an additional codebook for the groups induced by $g$. Based on this structure, Infomap tries to find a coarse graining $g$ that minimizes the resulting expected code length measured by the map equation~\cite[Sec.~3]{Rosvall_MapEquation}:
\begin{subequations}
 \begin{equation}\label{eq:map}
  \min_{g}\quad  q_{\curvearrowleft} \ent{\set{Q}} + \sum_{i=1}^{|\set{Y}|} p_{\circlearrowright}^i \ent{\set{P}^i}
 \end{equation}
 where
 \begin{equation}\label{eq:index}
  \ent{\set{Q}} = - \sum_{i=1}^{|\set{Y}|} \frac{q_{i\curvearrowleft}}{q_{\curvearrowleft}} \log \frac{q_{i\curvearrowleft}}{q_{\curvearrowleft}}
 \end{equation}
 and
 \begin{equation}\label{eq:modules}
  \ent{\set{P}^i} = -\frac{q_{i\curvearrowright}}{p_{\circlearrowright}^i} \log \frac{q_{i\curvearrowright}}{p_{\circlearrowright}^i} - \sum_{\alpha\in i} \frac{p_\alpha}{p_{\circlearrowright}^i} \log \frac{p_\alpha}{p_{\circlearrowright}^i}.
 \end{equation}
\end{subequations}
with $q_i=\Prob{Y_t=i}$, $q_{i\curvearrowright}=\Prob{Y_t=i,Y_{t+1}\neq i}$, $q_{i\curvearrowleft}=\Prob{Y_t\neq i,Y_{t+1}= i}$, $q_{\curvearrowleft}=\Prob{Y_{t+1}\neq Y_t}$, and $p_{\circlearrowright}^i = q_{i\curvearrowright} + q_i$. The coarse graining $g$ is thus only informative about $\stoch{X}$ (in the sense of~\eqref{eq:coarsegraining:symbol}) via representing a structural hierarchy in $\stoch{X}$ that admits an efficient encoding of realizations of $\stoch{X}$. Infomap successfully discovers this hierarchical structure, which justifies its use in community detection.

The authors of~\cite{Niedbalski_Aggregation} investigate coarse graining of absorbing Markov chains, i.e., of Markov chains that have states $\ell$ such that $\tpm_{\ell,\ell}=1$. In this case, the uncertainty of the trajectory from the initial state $m$ to absorption, measured by the conditional entropy $\lim_{T\to\infty} \ent{X_2^T|X_1=m}$, is finite. Modeling the evacuation behavior in large buildings, the authors of~\cite{Niedbalski_Aggregation} proposed putting states with similar corresponding entropies into the same group.

Lindqvist investigated the scenario where the quantity of interest is the initial state $X_1$ of the Markov chain, and the observation used to infer this state is a coarse graining of $X_t$ at some time $t$~\cite{Lindqvist_Loss}. In other words, Lindqvist is interested in the loss of information about $X_1$ induced by coarse graining $X_t$ to $g(X_t)=Y_t$. A natural information-theoretic formulation of this problem would be to consider the \emph{relevant information loss} $\mutinf{X_t,X_1}-\mutinf{Y_t;X_1}$. Instead, Lindqvist measured this loss of information via the following matrix norm\
\begin{equation}\label{eq:Lindqvist}
 \inf_M \Vert \tpm^t V M - \tpm^t \Vert
\end{equation}
where $\Vert A \Vert= \sup_i \sum_j |A_{i,j}|$, where $V$ is the coarse graining matrix introduced in~\eqref{eq:coarsegrainingmatrix}, and where the infimum is taken over all $|\set{Y}|\times|\set{X}|$ matrices. For an irreducible and aperiodic Markov chain $\stoch{X}$, as considered in this work,~\eqref{eq:Lindqvist} approaches zero as $t\to\infty$~\cite[Th.~3.2]{Lindqvist_Loss}, i.e., asymptotically $Y_t$ contains the same information about $X_0$ as $X_t$ (namely, no information at all). 

\subsection{Symbol-Wise Coarse Graining with Observations}\label{sec:coarsegraining:observation}

According to Problem~\ref{prob:coarsegraining}, this section is concerned with finding a coarse graining of a Markov chain $\stoch{X}$ that is informative about an observation process $\stoch{O}$. While we consider this general problem in Section~\ref{sec:coarsegraining:observation:plain}, we instantiate the problem differently in Section~\ref{sec:coarsegraining:observation:inference}. Namely, assuming that $(\stoch{X},\stoch{O})$ is a HMM and thus Markov, one can apply Problem~\ref{prob:coarsegraining} to $(\stoch{X},\stoch{O})$ rather than to $\stoch{X}$.

\subsubsection{Coarse Grainings that are Informative about an Observation Process}
\label{sec:coarsegraining:observation:plain}
On a very general level, this problem has been considered by Wolpert et al.~\cite{Wolpert_HighLevel}. Under the assumption that $\pmf{\stoch{O}|\stoch{X}}$ factorizes, i.e., that $(\stoch{X},\stoch{O})$ is a HMM, they aimed for a factorized map $\pmf{\stoch{Y}|\stoch{X}}(\stoch{y}|\stoch{x})=\prod_{t\in\naturals} \pmf{Y|X}(y_t|x_t)$ such that the future of $\stoch{O}$ is predicted with high accuracy from the current value of $\stoch{Y}$. The authors were seeking a reconstruction map $\pmf{\hat{\stoch{O}}|\stoch{Y}}$ such that $\hat{O}_{t+T}$ and $O_{t+T}$, $T\ge 0$, are similar in a well-defined sense, given observations of $\stoch{Y}$ up to time $t$. While they also consider exogenous similarity measures~\cite[Sec.~5.1]{Wolpert_HighLevel}, they argue in favor of information-theoretic costs in case no such similarity measure is provided. For example, the authors suggested to find a maximizer of~\cite[eq.~(4)]{Wolpert_HighLevel}
\begin{equation}\label{eq:wolpert}
 \max_{\pmf{Y|X}\in\set{P}_M} \max_{\pmf{\hat{O}|Y}} \mutinf{\hat{O}_{t+T};O_{t+T}}.
\end{equation}

Due to the Markov condition $O_{t+T}-X_{t+T} - X_t - Y_t - Y_{t+T} - \hat{O}_{t+T}$ and by the data processing inequality~\cite[Th.~2.8.1]{Cover_Information}, the inner maximization in~\eqref{eq:wolpert} cannot exceed $\mutinf{O_{t+T};Y_t}$, thus one may replace~\eqref{eq:wolpert} by
\begin{equation}\label{eq:wolpert:IB}
 \max_{\pmf{Y|X}\in\set{P}_M} \mutinf{O_{t+T};Y_t}.
\end{equation}
Replacing the fixed cardinality of $\set{Y}$ with a constrained capacity of the map $\pmf{Y|X}$ returns the Lagrangian formulation of Problem~\ref{prob:coarsegraining} by Lamarche-Perrin et al., who discussed state space reduction for agent-based models and proposed finding a minimizer of~\cite{Lamarche-Perrin_ABM}
\begin{equation}\label{eq:lamarche}
 \min_{\pmf{Y|X}} \mutinf{X_t;Y_t} - \beta \mutinf{O_{t+T}; Y_t}
\end{equation}
where $\beta>0$ trades between predictive power and the complexity of the reduced process $Y$. 

Both~\eqref{eq:wolpert:IB} and~\eqref{eq:lamarche} are instances of the information bottleneck (IB) problem~\cite{Tishby_InformationBottleneck}, with cardinality and rate constraints, respectively. The former problem is NP-hard~\cite{Mumey_NP}; however, it has an optimizer $\pmf{Y|X}$ that is a deterministic function $g{:}\ \set{X}\to\set{Y}$~\cite[Th.~1]{GeigerAmjad_HardClusters}, which provides justification to certain algorithmic approaches, such as~\cite{Slonim_Agglomerative,Slonim_DocClustering}. The latter problem is typically solved by iterating a set of equations until convergence, cf.~\cite{Tishby_InformationBottleneck}. Despite the complexity of the problem, certain settings allow simplifications. For example, under assumptions on the agent-based model and the deterministic map $g$, Lamarche-Perrin et al.\ showed that a further refinement of $g$ can not reduce the cost in~\eqref{eq:lamarche}, cf.~\cite[Th.~4]{Lamarche-Perrin_ABM}.

It is noteworthy how~\eqref{eq:wolpert:IB} and~\eqref{eq:lamarche} consider temporal dependence only via a particular forecasting horizon $T$ (or via an average over forecasting horizons, cf.~\cite[Sec.~5.2.2]{Wolpert_HighLevel}). I.e., these previous approaches consider the informativeness of the coarse graining of one time instance about the quantity of interest at the same or another time instance. As an alternative, one may measure this informativeness using mutual information \emph{rates}. Namely, we may seek a maximizer of\footnote{Equivalently, one may seek to minimize the conditional entropy rate $\entrate{\stoch{O}|\stoch{Y}}=\lim_{T\to\infty} \ent{O_1^T|Y_1^T}/T$.}
\begin{equation}\label{eq:coarsegraining:informationrate}
 \max_{\pmf{\stoch{Y}|\stoch{X}}} \mutrate{\stoch{Y};\stoch{O}} = \max_{\pmf{\stoch{Y}|\stoch{X}}} \lim_{T\to\infty} \frac{1}{T}\mutinf{Y_1^T;O_1^T}
\end{equation}
given that the limit exists. It is easy to imagine that~\eqref{eq:coarsegraining:informationrate} is considerably more difficult than the previous incarnations of Problem~\ref{prob:coarsegraining}, and that simple solutions are only possible in few and selected cases.

\subsubsection{Coarse Graining an Observation Process for State Inference}
\label{sec:coarsegraining:observation:inference}
We shall briefly stick to the setting of a HMM, but propose a different operational goal. Specifically, let $(\stoch{X}',\stoch{O}')$ be a HMM, i.e., $\stoch{X}'$ is Markov and $\pmf{\stoch{O}'|\stoch{X}'}$ factorizes, and let $\stoch{X}\equiv(\stoch{X}',\stoch{O}')$ and $\stoch{O}=\stoch{X}'$. Let us further restrict the mapping $\pmf{\stoch{Y}|\stoch{X}}$ to factorize as a deterministic function $g{:}\ \set{X}'\times\set{O}'\to\set{Y}$ that operates only on the observation process, i.e., $g(x',o')=g'(o')$ for some $g'{:}\ \set{O}'\to\set{Y}$. Thus, we consider the scenario in which an observation $\stoch{O}'$ is coarse grained with the aim of being most informative about $\stoch{X}'$. Such a problem is relevant, for example, in receiver design, where $\stoch{X}'$ approximates a sequence of transmitted code words, $\stoch{O}'$ represents its observations at the receiver input affected by a memoryless channel, and the subsequent coarse graining is implemented by a quantizer that transforms the (often continuous-valued) receiver input signal to a signal that a digital architecture can work with. For example, the authors of~\cite{Lewandowsky_ReceiverDesign} determined a minimizer of
\begin{equation}\label{eq:lewandowsky}
 \min_{g'\in\set{P}_M} \mutinf{O'_t;Y_t}
\end{equation}
for implementing discrete message passing decoders for LDPC codes. While in general~\eqref{eq:lewandowsky} is NP-hard, it can be shown that for $\stoch{X}$ having a binary state space $\set{X}=\{0,1\}$, the above optimization problem can be solved optimally in $\set{O}(|\set{O}'|^3)$~\cite{Kurkoski_Quantization}.

\section{Information-Theoretic Approaches to Markov Chain Model Reduction}\label{sec:modelreduction}
In this section we will investigate several information-theoretic approaches to model reduction. We start with reducing the state space of a Markov chain in Section~\ref{sec:modelreduction:aggregation}, drawing a connection to coarse graining discussed in Section~\ref{sec:coarsegraining}. We then review approaches that preserve the state space, but reduce the number or complexity of model parameters in Section~\ref{sec:modelreduction:parameters}. We finally discuss approaches that are related to the compressibility of the resulting Markov chain $\tilde{\stoch{X}}$ (Section~\ref{sec:modelreduction:compressibility}).

\subsection{Markov Aggregation}\label{sec:modelreduction:aggregation}
Markov aggregation refers to the task of replacing a Markov chain $\stoch{X}$ on the large state space $\set{X}$ by a Markov chain $\tilde{\stoch{X}}$ on a significantly smaller state space $\tilde{\set{X}}$, under the requirement that the aggregated chain $\tilde{\stoch{X}}$ is similar -- in a well-defined sense -- to $\stoch{X}$. Very often, such an aggregation is obtained by clustering the states $\set{X}$ of the original Markov chain. Letting $g{:}\ \set{X}\to \tilde{\set{X}}$ denote the clustering function, the intimate relationship between Markov aggregation and Markov coarse graining becomes clear. Indeed, even the terms ``aggregation'' and ``coarse graining'' are often used synonymously in the literature.

\subsubsection{Coarse Graining-Based Markov Aggregation}

Given a fixed coarse graining $g$, a popular approach is to define the transistion probability matrix $\agg=[\agg_{i,j}]_{i,j\in\tilde{\set{X}}}$ of $\tilde{\stoch{X}}$ by the one-step conditional probabilities of the coarse-grained process $\stoch{Y}$, i.e., by\footnote{Matrix formulations of this statement can be found in~\cite{Amjad_GeneralizedMA,GeigerEtAl_OptimalMarkovAggregation} and are derived from the matrix notation in~\cite[\S6.3]{Kemeny_FMC}.}
\begin{equation}\label{eq:aggmat}
 \agg_{i,j} = \Prob{Y_2=j|Y_1=i} = \frac{\sum_{\ell\in g^{-1}(i)}\sum_{m\in g^{-1}(j)} \pi_\ell \tpm_{\ell,m}}{\sum_{\ell\in g^{-1}(i)} \pi_\ell}
\end{equation}
where $\pi$ is the unique invariant distribution of $\stoch{X}$ (we assumed that all Markov chains are irreducible and aperiodic).\footnote{For absorbing Markov chains, the invariant distibution is trivial or not unique, depending whether there is just a single or multiple absorbing states, respectively. In this case, $\pi$ in~\eqref{eq:aggmat} must be replaced by different weights. Examples include uniform weights or weights derived from the potential matrix of the absorbing Markov chain, cf.~\cite[(20) \& (23)]{Niedbalski_Aggregation}.} This choice of $\agg$ is justified by the fact that the resulting aggregated process $\tilde{\stoch{X}}$ has minimal Kullback-Leibler divergence rate (KLDR) to the coarse-grained process $\stoch{Y}$, cf.~\cite[Lemma~3]{GeigerEtAl_OptimalMarkovAggregation} or~\cite[Cor.~10.4]{Gray_Entropy}.

These considerations yield, for a given coarse graining, the aggregation optimal in the sense of the KLDR. The Markov aggregation problem thus requires solving the coarse graining problem. In turn, all approaches to coarse graining discussed in Section~\ref{sec:coarsegraining} are valid approaches to Markov aggregation as well, where the objectives for coarse graining determine the similarity measure $\mathsf{Sim}(\cdot\to\cdot)$ in~\eqref{eq:modelreduction}.

A different approach to Markov aggregation is to directly prescribe a similarity measure $\mathsf{Sim}(\stoch{X}\to\tilde{\stoch{X}})$. This approach was taken, for example, by the authors of~\cite{Meyn_MarkovAggregation,Deng_ITFramework}, who proposed to \emph{lift} the aggregated chain $\tilde{\stoch{X}}$ to the state space $\set{X}$ and to then measure similarity via the KLDR between the original and the lifted Markov chains $\stoch{X}$ and $\tilde{\stoch{X}}'$. Specifically, the authors proposed finding a minimizer of
\begin{equation}\label{eq:minKLDR}
 \min_{g\in \set{P}_M} \kld{\stoch{X}}{\tilde{\stoch{X}}'}
\end{equation}
where
\begin{equation}\label{eq:KLDR}
\kld{\stoch{X}}{\tilde{\stoch{X}}'} = \sum_{\ell,m\in\set{X}} \pi_\ell \tpm_{\ell,m} \log\frac{\tpm_{\ell,m}}{\tpm_{\ell,m}'}
\end{equation}
is the KLDR between two Markov chains on the same state space, cf.~\cite{Rached_KLDR}, and where
\begin{equation}\label{eq:pi-lifting}
 \tpm_{\ell,m}' = \frac{\pi_\ell}{\sum_{l\in g^{-1}(g(\ell))}\pi_l} \agg_{g(\ell),g(m)}
\end{equation}
is the $\pi$-lifting of $\agg$~\cite[Def.~2]{Meyn_MarkovAggregation} (see also~\cite{Vidyasagar_DifferentCardinalities}). Remarkably, it can be shown that the coarse graining $g$ that minimizes~\eqref{eq:minKLDR} is exactly the coarse graining that maximizes $\mutinf{Y_1;Y_2}$~\cite[Lemma~3]{Meyn_MarkovAggregation}. Thus, maximizing the temporal dependence structure subject to a cardinality constraint on the coarse-grained process' state space makes the lifting of the aggregated Markov chain similar to the original Markov chain. This lifting was also used in~\cite{Petrov_Reaction} for aggregating continuous-time Markov chains for combinatorial chemical reaction systems.

Geiger et al.\ replaced the lifting in~\eqref{eq:pi-lifting} by one that depends on the original Markov chain's transition probability matrix $\tpm$, i.e., they set~\cite[Def.~7]{GeigerEtAl_OptimalMarkovAggregation}
\begin{equation}
 \tpm_{\ell,m}' =\begin{cases}
                   \frac{\tpm_{\ell,m}}{\sum_{k\in g^{-1}(g(m))}\tpm_{\ell,k}} \agg_{g(\ell),g(m)}, & \sum\limits_{k\in g^{-1}(g(m))}\tpm_{\ell,k}>0\\
                   \frac{1}{|g^{-1}(g(m))|}\agg_{g(\ell),g(m)},& \text{else.}
                  \end{cases}
\end{equation}
As a consequence of this lifting, the resulting KLDR simplifies to $\ent{Y_2|Y_1}-\ent{Y_2|X_1}$, cf.~\eqref{eq:lumpability_optimization} or~\cite[(31)--(34)]{GeigerEtAl_OptimalMarkovAggregation}. Thus, the coarse graining $g$ minimizing~\eqref{eq:minKLDR} makes $\stoch{Y}$ ``as Markov as possible''. Such liftings were only presented for the coarse graining objectives in~\cite{Meyn_MarkovAggregation} and \cite{GeigerEtAl_OptimalMarkovAggregation}; it was not possible to construct corresponding liftings for coarse grainings to higher-order Markov chains~\cite{GeigerWu_HigherOrder}, nor for the generalized framework presented in~\cite{Amjad_GeneralizedMA} that trades between the objectives of~\cite{Meyn_MarkovAggregation,GeigerEtAl_OptimalMarkovAggregation}.

Lifting a Markov chain from a smaller state space to one on a larger state space is one way to use standard information-theoretic quantities, such as the KLDR, as a similarity measure $\mathsf{Sim}(\stoch{X}\to\tilde{\stoch{X}})$. A different approach is to define an appropriate similarity measure that directly accepts Markov chains on different state spaces. Such an approach was taken by~\cite{Xu_Reduction,Xu_ReductionConf,Sledge_VoI}. Considering again Markov aggregation via coarse graining, the authors proposed quantities depending on the joint process $(\stoch{X},\tilde{\stoch{X}})$ as optimization objectives. For example, and instantiated to Markov chains, the authors of~\cite{Xu_Reduction,Xu_ReductionConf} proposed to minimize the expected Kullback-Leibler divergence between the outgoing transition probabilities of a given state and the probabilities connecting states of $\tilde{\stoch{X}}$ with states of $\stoch{X}$. Effectively, the authors proposed minimizing
\begin{equation}\label{eq:XU_cost}
\mutinf{X_1;X_2}  - \mutinf{X_2;Y_1} -\gamma \ent{Y_2|X_1}
\end{equation}
where $\gamma$ is an annealing parameter that is reduced in subsequent optimization iterations. Similarly, the authors of~\cite{Sledge_VoI} rely on a model for the joint process $(\stoch{X},\tilde{\stoch{X}})$ and measure similarity via a modified Kullback-Leibler divergence and the value of information, respectively.

The above works require, as input, the size of the aggregated state space $\tilde{\set{X}}$. Determining an appropriate size is a difficult problem, especially if the Markov chain exhibits structure on several hierarchies. Nevertheless, the authors of~\cite{Meyn_MarkovAggregation} and~\cite{GeigerEtAl_OptimalMarkovAggregation} proposed determining $|\tilde{\set{X}}|$ by looking for ``elbows'' or minima in their respective objective functions. For the approach in~\cite{Xu_Reduction,Xu_ReductionConf}, the size of the aggregated state space was connected to the heterogeneity of aggregate states~\cite{Xu_Number}. The authors of~\cite{Sledge_Number}, in turn, proposed to modify the value-of-information criterion to prevent the aggregated state space from becoming too large.

The approach in~\cite{Piccardi_CD} directly optimizes over the size of the aggregated state space. With an application in community detection, the authors suggest to find a function $g$ from $\set{X}$ into a set $[M]$ as large as possible so that the persistence probabilities $\agg_{i,i}$, $i\in [M]$ of the resulting Markov aggregation all exceed a certain predefined value $\alpha$. I.e., the authors aim at finding coarse grainings $g$ maximizing
\begin{equation}\label{eq:piccardi}
 \max_M\quad \{g{:}\ \set{X}\to[M]{:}\ \forall i\in[M]{:}\ \agg_{i,i}\ge \alpha\}
\end{equation}
where $\agg$ is given in~\eqref{eq:aggmat}. From an algorithmic perspective, the coarse graining $g$ is obtained by clustering states based on the similarity between their $T$-step transition probabilities $\tpm^T$. In other words, while the intended cost function~\eqref{eq:piccardi} does not admit an immediate interpretation in terms of the similarity formulation of Problem~\ref{prob:modelreduction}, its algorithmic implementation admits such an interpretation.

\subsubsection{Aggregation of HMMs}\label{sec:HMM}
As mentioned in Section~\ref{sec:overview}, model reduction of Markov chains includes model reduction of HMMs $(\stoch{X}',\stoch{O})$ as special case. However, in some settings it may prove beneficial to exploit the special structure of an HMM when its state or observation space shall be aggregated. Aggregation of HMMs can happen via coarse graining or otherwise. Suppose that $\stoch{Y}_X$ is a coarse graining of the state process $\stoch{X}'$ and that $\stoch{Y}_O$ is a coarse graining of the state process $\stoch{O}$. Then, the result from~\eqref{eq:aggmat} can be carried over to HMMs:
\begin{thm}\label{thm:hmmaggregation}
 Let $\stoch{Y}=(\stoch{Y}_X,\stoch{Y}_O)$ be a stationary process on $\set{X}\times\set{O}$. The hidden Markov model $\tilde{\stoch{X}}=(\tilde{\stoch{X}}',\tilde{\stoch{O}})$ minimizing the KLDR to $\stoch{Y}$ satisfies
 \begin{subequations}
 \begin{align}
   \tilde\tpm_{\ell,m} &=\pmf{\tilde X'_t|\tilde X'_{t-1}}(m|\ell) = \pmf{{Y}_{X,t}|{Y}_{X,t-1}}(m|\ell)\\
   W_{\ell,i} &= \pmf{\tilde O_t|\tilde X'_t}(i|\ell) = \pmf{Y_{O,t}|Y_{X;t}}(i|\ell)
 \end{align}
 \end{subequations}
\end{thm}

\begin{IEEEproof}
 See Appendix~\ref{proof:hmmaggregation}.
\end{IEEEproof}

The same approximation as in Theorem~\ref{thm:hmmaggregation} was also proposed in~\cite[Sec.~IV.C]{Vidyasagar_MarkovAgg}, albeit without justification via the KLDR. Further, this approximation also minimizes the KLDR between the original HMM and a lifting of the aggregated HMM~\cite[Th.~2]{Deng_HMM}.

The authors of~\cite{Deng_HMM} generalized the lifting~\eqref{eq:pi-lifting}  from~\cite{Deng_ITFramework,Meyn_MarkovAggregation}, to the aggregation of HMMs. Specifically, the authors focused on aggregating the state process $\stoch{X}'$ using a coarse graining $g$, leaving the state space of the observation process $\stoch{O}$ unchanged. Utilizing the lifting and measuring similarity using the KLDR $\kld{\stoch{X}',\stoch{O}}{\tilde{\stoch{X}}',\tilde{\stoch{O}}}$, they arrived at the following optimization problem~\cite[Th.~3]{Deng_HMM}:
\begin{equation}\label{eq:deng_HMM}
 \min_{g{:}\ \set{X}\to[M]} \mutinf{X_1;X_2} - \mutinf{Y_1;Y_2} + \mutinf{O_1;X_1|Y_1}
\end{equation}
While the KLDR $\kld{\stoch{X}',\stoch{O}}{\tilde{\stoch{X}}',\tilde{\stoch{O}}}$ in this case still allows for a closed-form expression for $(\tilde{\stoch{X}}',\tilde{\stoch{O}})$ derived from Theorem~\ref{thm:hmmaggregation}, the same does not hold for the KLDR between the observation processes $\kld{\stoch{O}}{\tilde{\stoch{O}}}$. For this setting, the authors of~\cite{Deng_Recursive} propose an approach that recursively optimizes a stochastic (or ``soft'') coarse graining.

Regarding classical similarity measures, Kotsalis et al.~\cite{Kotsalis_HMMReduction} solve the HMM agreggation problem as a two-step problem. Their approach is based on the fact that an HMM is a special case of a generalized automaton, and that generalized automata are equivalent to a certain class of jump linear systems. The state space of these latter systems can be reduced such that the resulting error is bounded. In the first step of the proposed approach, this equivalence is used to coarse grain the state process $\stoch{X}'$ of the HMM while ensuring that the induced squared error on the observation process $\stoch{O}$ is bounded. Since the reduced generalized automaton may not be a HMM, in the second step a non-convex optimization problem is solved that finds a reduced HMM with state space $\tilde{\set{X}}'$ such that this error is minimized. This ensures that the aggregated HMM exhibits similar probabilistic properties in the sense that $\sum_{o^*\in\set{O}^*} \left(\pmf{{\tilde{O}}^*}(o^*)-\pmf{O^*}(o^*)\right)^2$ is small, where $\pmf{O*}$ and $\pmf{\tilde{O}^*}$ are the PMFs of arbitrary-length sequences of $\stoch{O}$ and $\tilde{\stoch{O}}$, respectively. Thus, similarity in the sense of Problem~\ref{prob:modelreduction} is here measured by the quadratic divergence between process distributions. Another example for a classical similarity measure is given in~\cite{Wu_HMMCompression}, where the authors propose to approximate a HMM by a different HMM such that the (finite-length) observation sequences are similar in the sense of the total variation distance $\Vert \pmf{O_1^T} - \pmf{\tilde{O}_1^T}\Vert$. 

\subsubsection{Other Approaches to Markov Aggregation}
In~\cite{Tzortzis_TVD}, similarity between the original and the aggregated Markov chain was measured by the total variation distance between the respective invariant distributions. More specifically, the authors proposed to find a distribution vector $\nu$ on the state space $\set{X}$ that has maximum entropy, under the constraint that  the total variation distance to the invariant distribution $\pi$ does not exceed $R$~\cite[Problem~2.9]{Tzortzis_TVD}. For increasing $R$, the authors observed that the approximating vector $\nu$ consists of increasingly large groups of identical entries~\cite[Table~III]{Tzortzis_TVD}, based on which a coarse graining and, subsequently, an aggregated Markov chain with transition probability matrix $\agg$ as in~\eqref{eq:aggmat} can be constructed.

A special case of Markov aggregation was considered in~\cite{Peixoto_CD}. For a Markov chain with order $k$ and a $|\set{X}|^k\times|\set{X}|$ transition probability matrix $\tpm'$, they assume a model of the type $\tpm=V' \agg' U$, where $V'$ is a coarse graining matrix mapping $k$-sequences of previous states to $M'$ state sequence clusters, $\agg'$ is the transition probability matrix between these $M'$ state sequence clusters and $M$ state clusters, and where $U$ contains the probabilities from state clusters to states~\cite[eq.~(5)]{Peixoto_CD}. Thus, the authors propose a \emph{co-clustering} of states and state $k$-sequences of the original Markov chain $\stoch{X}$. The matrices $V'$, $U$, and $\agg'$ are obtained by minimizing the description length of a Bayesian generative model, thus trading between the number of co-clusters $M'$, $M$ and the Markov order $k$ on the one side and the quality of fit on the other side.

\subsection{Simplifying Model Parameterization}\label{sec:modelreduction:parameters}

A different approach to Markov model reduction is to reduce the number or complexity of its model parameters. In this case, the model complexity can be reduced while still maintaining the original state space, i.e., in general we have $\tilde{\set{X}}=\set{X}$. One approach in this direction, especially for higher-order Markov chains, is to parameterize the transition probabilities $\tpm_{\ell,m}$ via function approximators, such as neural networks. In this section, however, we will focus on more classical approaches. This can mean replacing the transition probability matrix by a low-rank approximation, storing its entries with finite precision, or removing transitions that occur with low probability.

\subsubsection{Low-Rank Approximation}
The authors of~\cite{Ghasemi_NMF} seek a non-negative matrix tri-factorization of $\tpm\approx W \agg U=:\tilde\tpm$, i.e., they seek stochastic matrices $U$, $W$, and $\agg$, where $\agg$ is $M\times M$, $M< |\set{X}|$. If such a factorization exists, then the $T$-step transition probabilities $\Prob{X_{t+T}=x'|X_t=x}$ can be computed in $\mathcal{O}(TM^2)$ instead of $\mathcal{O}(T|\set{X}|^2)$~\cite[Prop.~2]{Ghasemi_NMF}, as would be required to evaluate the $T$-fold matrix product $\tpm^T$. 
The authors aim for sparse aggregation and deaggregation maps $W$ and $U$, respectively. Since enforcing sparsity by constraining the $\ell_0$-norm of the rows of $W$ and $U$ is NP-hard, the authors suggest to rather seek minimizers of the following optimization problem:
\begin{equation}\label{eq:ghasemi}
  \min_{W, \agg, U} \frac{1}{2} \Vert \tpm - W\agg U\Vert_F^2 +\lambda_u \Vert U\Vert_1 + \lambda_w \Vert W\Vert_1
\end{equation}
where $\lambda_u,\lambda_w>0$ are regularization parameters, where $\Vert\cdot\Vert_F$ is the Frobenius norm, where $\Vert\cdot\Vert_1$ is the $\ell_1$-norm, and where the minimization is performed over the sets of stochastic matrices with dimensions $|\set{X}|\times M$, $M\times M$, and $M\times |\set{X}|$, respectively. Note that in~\eqref{eq:ghasemi}, the Frobenius norm measures the similarity between $\stoch{X}$ and $\tilde{\stoch{X}}$ and the optimization is constrained via fixing the rank $M$. The authors propose a gradient-based approach to solve the minimization problem and show that it converges under certain conditions on the step size parameters~\cite[Th.~1]{Ghasemi_NMF}. Rather than fixing the rank $M$ of the reduced model, the authors of~\cite{Duan_LowRank} propose to instead regularize the non-negative rank. Relaxing the non-negative rank to a regularizer $\Omega(\cdot)$ based on the atomic norm, the convex optimization problem reads
\begin{equation}\label{eq:duan}
 \min_{\tilde \tpm} \frac{1}{2} \Vert \tpm - \tilde\tpm\Vert_F^2 +\lambda \Omega(\tilde\tpm)
\end{equation}
where it is assumed that $\tilde\tpm$ thus obtained has a non-negative factorization $\tilde\tpm=W\agg U$. 

While the work of~\cite{Ghasemi_NMF} resembles Markov aggregation, the authors of~\cite{Deng_LowRank} did not enforce a tri-factorization structure on their low-rank approximation. Specifically, the authors proposed to minimize the KLDR $\kld{\stoch{X}}{\tilde{\stoch{X}}}$ between the original and the reduced Markov chain. Since a constraint on the rank would render this a non-convex optimization problem, the authors rather propose to regularize the optimization via the nuclear norm\footnote{Note that the nuclear norm does not guarantee a non-negative factorization of $\tilde\tpm$, while the atomic norm-based regularizer of~\cite{Duan_LowRank} does.} of the spectral decomposition of $\tilde\tpm$. Specifically, they aim to find a minimizer of~\cite[eq.~(5)]{Deng_LowRank}
\begin{equation}
 \min_{\tilde\tpm} \kld{\stoch{X}}{\tilde{\stoch{X}}} + \beta \Vert S(\tilde\tpm)\Vert_*
\end{equation}
where $\tilde{\stoch{X}}$ has transition probability matrix $\tilde\tpm$, $S(A)$ is the stochastic decomposition of the reversible transition matrix $A$, and $\Vert A \Vert_*$ is the nuclear norm of $A$, i.e., the sum of all of $A$'s singular values.

\subsubsection{Quantizing or Pruning the Transition Probability Matrix}
Model parametrization can also be simplified by directly operating on the transition probability matrix $\tpm$. For example, a simple baseline approach is to remove all transitions that occur with a probability lower than a given threshold $\tau$, i.e.,~\cite[eq.~(33)]{Henter_EntropyRate}
\begin{equation}
 \hat\tpm_{\ell,m} \propto \begin{cases}
                             \tpm_{\ell,m}, & \tpm_{\ell,m}\ge \tau \\ 0, & \text{else}.
                            \end{cases}
\end{equation}
where $\tilde\tpm$ is eventually obtained normalizing the row sums of $\hat\tpm$ to 1. Slightly less ad-hoc is the quantization procedure proposed in~\cite[Sec.~IV]{GeigerBoecherer_GreedyQuant_arXiv}. There, the authors proposed to quantize every row of $\tpm$ with a quantization bin size $1/B$, for which they present a greedy algorithm that is optimal in the sense of the Kullback-Leibler divergence between the original row vector and its quantization. As an immediate consequence, this procedure yields a quantized transition probability matrix with precision $1/B$ that minimizes the KLDR $\kld{\stoch{X}}{\tilde{\stoch{X}}}$. The resulting approximate model $\tilde{\stoch{X}}$ has the same transition graph as $\stoch{X}$, i.e., no transitions are removed~\cite[Lemma~1]{GeigerBoecherer_GreedyQuant_arXiv}. If it is desirable that edges are removed, then instead individual rows can be quantized according to the total variation distance or the reversed Kullback-Leibler divergence; see~\cite{Geiger_OptimalQuantization} for optimal algorithms.\footnote{Taking the reversed Kullback-Leibler divergence, for example, makes it more difficult to evaluate the KLDR as a similarity measure. Indeed, $\mathsf{Sim}(\stoch{X}\to\tilde{\stoch{X}})=\kld{\tilde{\stoch{X}}}{\stoch{X}}$ depends on the invariant distribution of $\tilde{\stoch{X}}$ (cf.~\eqref{eq:KLDR}), which is not straightforward to compute analytically.}

Related to pruning is the idea of removing states with short persistence times. Jia did this for irreversible Markov chains in~\cite{Jia_Removal} and computed the resulting reduced transition probability matrix $\tilde\tpm$. More specifically, since the authors focused on continuous-time Markov chains, they removed states with fast leaving rates and showed that the resulting reduced model exhibits similar behavior as the slow states of the original chain.

The authors of~\cite{Tzortzis_TVD} propose to remove states of a Markov chain by optimizing the observation matrix of a HMM. Specifically, for $W=[W_{\ell,i}]=\Prob{Y_t=i|X_t=\ell}$ and if $Y_t\in\set{Y}\subseteq\set{X}$, the authors propose maximizing~\cite[Problem~2.3]{Tzortzis_TVD}
\begin{equation}
  \sum_{\ell\in\set{X}} \sum_{i\in\set{Y}} \pi_\ell W_{\ell,i} c_i
\end{equation}
for a vector $c$ of rewards and under the condition that the total variation distance between $\tpm_{\ell,\cdot}$ and the extension of $W_{\ell,\cdot}$ to $\set{X}$ does not exceed $R$. For increasing $R$, more and more columns of $W$ become zero, based on which the authors propose to delete the corresponding states. Similarly, the authors arrive at the removal of states by a criterion involving only an approximation of the invariant distribution $\pi$~\cite[Problem~2.8]{Tzortzis_TVD}.

\subsubsection{Other Approaches}
A problem similar to the one in~\cite{Ghasemi_NMF} was considered in~\cite{Georgiou_MarkovCreditRisk}, where the authors searched for an approximation $\tilde\tpm$ that is lumpable for a \emph{given} coarse graining $g$. With $V$ as in~\eqref{eq:coarsegrainingmatrix} and  $U$ being the Moore-Penrose pseudo-inverse of $V$, the authors seek a minimizer of the following problem:
\begin{align}
 &\min_{\tilde\tpm} \Vert \tpm-\tilde\tpm\Vert_2\\
 \text{s.t.} \quad& VU\tilde\tpm V = \tilde\tpm V\notag\\
 & \tilde\tpm \text{ is row stochastic.}\notag
\end{align}
(The condition $VU\tilde\tpm V = \tilde\tpm V$ is a sufficient condition for lumpability~\cite[Th.~6.3.5]{Kemeny_FMC}.) The authors propose solving this problem using Dykstra's algorithm. The main difference between~\cite{Ghasemi_NMF} and~\cite{Georgiou_MarkovCreditRisk} is that the former also optimizes over the (stochastic) coarse graining $W$ (or $V$), while in the latter the coarse graining matrix $V$ is given. Further, while the approximation $\tilde\tpm$ in~\cite{Ghasemi_NMF} has low rank, the approximation in~\cite{Georgiou_MarkovCreditRisk} may not be so.

The authors of~\cite{Toth_PAM} simplify the model parameterization by prescribing that $\tilde\tpm$ has a specific structure. Specifically, also assuming a candidate coarse graining $g$, the authors require that $\tilde\tpm$ is constructed as follows:
\begin{equation}
\tilde\tpm_{\ell, m} = 
\begin{cases}
r_m^{g(m)}  \cdot (1-s_{g(\ell)}), & g(\ell)=g(m),\\
r_m^{g(m)}  \cdot s_{g(\ell)}     \cdot\frac{u_{g(m)}}{1 - u_{g(\ell)}}, &\text{otherwise}
\end{cases}
\end{equation}
where $\{r^{i}\}_{i\in\set{Y}}$ are probability vectors (i.e., $r_\ell^{i}\ge 0$ and $\sum_{\ell\in g^{-1}(i)} r_\ell^{i}=1$), $u$ is a probability vector, and where $s_i\in [0,1]$. Thus, the reduced Markov chain switches state groups by a Bernoulli coin flip parameterized by $s_i$. Movements within state groups are independent and identically distributed, and movements between state groups are parameterized by a single probability vector $u$. The reduced transition probability matrix $\tilde\tpm$ hence consists of groups of identical rows and thus yields $\tilde{\stoch{X}}$ lumpable w.r.t.\ the coarse graining $g$.

The authors subsequently proposed selecting  $\{r^{i}\}$, $u$, and $\{s_i\}$ by minimizing the KLDR to the original Markov chain $\stoch{X}$, i.e., by finding minimizers to
\begin{equation}\label{eq:toth}
 \min_{\{r^{i}\},u,\{s_i\}} \kld{\stoch{X}}{\tilde{\stoch{X}}}.
\end{equation}
The minimization yields closed-form expressions $r_\ell^{i}=\Prob{X_t=\ell,Y_t=i}$ and $s_i=\Prob{Y_{t+1}\neq i|Y_t=i}$. Selecting $u_i=\Prob{Y_t=i}$ as sub-optimal choice simplifies the KLDR in~\eqref{eq:toth} and subsequently facilitates the search for a coarse graining $g$ minimizing this objective~\cite[eq.~(13) \& Prop.~1]{Toth_PAM}. The resulting optimization problem was proposed for community detection.

\subsection{Compressibility}\label{sec:modelreduction:compressibility}
Yet another goal for Markov model reduction is to ensure that the model generates easily compressible sequences of realizations. This may be particularly useful if the model is obtained from observing noisy or distorted data, and one aims to retrieve the underlying, noise-free model, assuming that noiseless realizations are more compressible than noisy ones. In this case, Problem~\ref{prob:modelreduction} is constrained by requiring that the reduced model $\tilde{\stoch{X}}$ has low entropy, description length, or any other measure of compressibility.

One noteworthy work in this direction is~\cite{Henter_EntropyRate}, where the authors proposed replacing a process $\stoch{X}$ by a reduced version $\tilde{\stoch{X}}$ that has a minimal entropy rate subject to a certain distortion criterion. Selecting the negative KLDR as similarity measure $\mathsf{Sim}(\stoch{X}\to\tilde{\stoch{X}})$, the authors seek a minimizer of
\begin{equation}\label{eq:henter}
 \min_{\tilde\tpm} \entrate{\tilde{\stoch{X}}} + \lambda \kld{\tilde{\stoch{X}}}{\stoch{X}}.
\end{equation}
Note that here the KLDR is computed w.r.t.\ to reduced model rather than w.r.t.\ the original model (cf.~\eqref{eq:minKLDR}). The reason for this reversal of roles is that in~\eqref{eq:minKLDR} the process $\stoch{X}$ is the original underlying model that is approximated by $\tilde{\stoch{X}}$. In contrast, here in~\eqref{eq:henter} the reduced model $\tilde{\stoch{X}}$ is assumed to be the true underlying model that should be inferred from the model $\stoch{X}$ inferred from noisy data. The authors show that~\eqref{eq:henter} allows closed-form solutions for Gaussian processes and Markov chains. Indeed, the optimizing transition probability matrix $\tilde\tpm$ is~\cite[eq.~(28)]{Henter_EntropyRate}
\begin{equation}
 \tilde\tpm = \frac{1}{\nu} \diag{\mu}^{-1} \tpm^{(\alpha)} \diag{\mu}
\end{equation}
where $\alpha=\lambda/(\lambda-1)$, where $\tpm^{(\alpha)}$ is the elementwise exponentiation of $\tpm$, and where $\mu$ is the unique positive right eigenvector of $\tpm^{(\alpha)}$ satisfying $\tpm^{(\alpha)}\mu = \nu\mu$.

\section{Discussion and Implications for Knowledge Discovery and Data Mining}
\label{sec:discussion}

After taking a closer look at the two problems of coarse graining and model reduction, they seem vastly different. Indeed, the object over which the coarse graining problem optimizes is always a conditional distribution (or deterministic function), so the approaches discussed in Section~\ref{sec:coarsegraining} differ either in the setup (symbol-wise or block-wise coarse graining, including or excluding a separate quantity of interest, etc.) or in the selected optimization objective (preserving process information, temporal dependence, preserving the Markov property, etc.). In contrast, for model reduction we have found in Section~\ref{sec:modelreduction} that most of the literature aims for a high similarity between the reduced and the original model (as quantified using the Kullback-Leibler divergence rate or a matrix norm of the difference between the original and the approximating transition probability matrices). The main differences between the model reduction approaches turned out to be caused by different modeling assumptions for the reduced-order models, such as reduced alphabets, low-rank or low-accuracy approximations of the original transition probability matrix, or aspects of compressibility.

There is a setting, however, where coarse graining and model reduction overlap, namely when the reduced model is based on a coarse graining (Section~\ref{sec:modelreduction:aggregation}). In this setting, we still aim to find a conditional distribution (or deterministic function), but take some similarity between the original Markov chain and the coarse-grained process as optimization objective (cf. Section~\ref{sec:overview:lumpability}). In this way, the probabilistic description of the coarse-grained process can step in as a model that is reduced in the sense of a (significantly) smaller state space. 

In addition to this setting, there is a certain grey area between model reduction and coarse graining, making it difficult to assign a certain method to either of these categories. This grey area results, to some extent, from the fact that the outcome of every optimization problem is equally influenced by the applied modeling assumptions, the optimization objective, and the optimization algorithm. For example, the approach of~\cite{Ghasemi_NMF} in~\eqref{eq:ghasemi} seeks a low-rank approximation of $\tpm$ that has an explicit structure. The \emph{modeling assumption} is that $\tpm\approx W\agg U$, where $W$, $\agg$, and $U$ are stochastic matrices. Indeed, if $W$ is defined by a deterministic coarse graining $g$, then the Markov chain with transition probability matrix $W\agg U$ is lumpable w.r.t.\ $g$. Thus, while the authors of~\cite{Ghasemi_NMF} seek a sparse low-rank approximation in the \emph{optimization objective}, their modeling assumption implies that the reduced model is at least approximately lumpable w.r.t.\ a specific coarse graining. This brings their general work close to both coarse graining-based Markov aggregation (Section~\ref{sec:modelreduction:aggregation}) and to coarse grainings that preserve the Markov property (Section~\ref{sec:coarsegraining:simple:lumpable}).

We finally take up the discussion that we initiated in the introduction, stating that model reduction methods for Markov chains can be used to solve unsupervised machine learning problems. This is most obvious for coarse graining or coarse graining-based Markov aggregation, as coarse grainings connect naturally with the concept of \emph{patterns} as abstractions of data. Information-theoretic approaches to Markov chain coarse graining or Markov aggregation can thus be used in the exploratory analysis of time series data. Indeed, the authors of~\cite{GeigerWu_HigherOrder} showed that a coarse graining $g$ that maximizes $\mutinf{Y_3;Y_1^2}$ for a letter bi-gram model derived from a famous text induces a meaningful partition of the English alphabet. Similarly, a coarse graining that preserves the temporal dependence structure $\mutinf{Y_2;Y_1}$ was used to divide complex hand movements into simple components for movement trajectories recorded from monkeys~\cite{Goldberger_Movements} and humans~\cite{Erez_Movements}.

In addition to exploring time series data, many other problems in unsupervised machine learning can be transformed into coarse graining problems of appropriately derived Markov chains (or, random walks). For example, community detection in graphs concerns finding a partition of the vertex set of a graph such that vertices within a given element of the partition share more edges with each other than with vertices in other elements of the partition. We can now identify the vertex set of the graph with the state space of a Markov chain and derive the Markov chain's transition probability matrix from the set of edges (or edge weights) of the graph. (The most obvious approach is to set $\tpm_{\ell,m}\propto w_{\ell,m}$, where $w_{\ell,m}$ is the weight of the edge $\ell\to m$, but many other possibilities exist.) Then, community detection, which is essentially the problem of clustering the vertex set, is transformed into coarse graining the thus derived Markov chain. Indeed, coarse graining-based approaches to community detection have been proposed in~\cite{Faccin_CD,Lambiotte_RandomWalks,Hurley2015,Hurley2016,Piccardi_CD,Rosvall_Infomap}. Similarly, other clustering problems can be transformed to Markov chain coarse graining problems by identifying the dataset with the state space and by deriving the transition probability matrix from pairwise (dis)similarities between data points, cf.~\cite{Steger_SemiSupervised,Alush_PairwiseClustering,Tishby_MarkovRelaxation}. Even co-clustering, the problem of simultaneously clustering two related datasets, can be identified as a coarse graining problem by considering a random walk on a bi-partite graph~\cite{BloechlEtAl_MCClustering}.

While the connection between clustering and coarse graining is easy to see, also other approaches to Markov model reduction can be used to solve unsupervised machine learning problems. As just one example we point at the community detection method proposed in~\cite{Toth_PAM}, where the authors aimed for a simplified parameterization of a Markov chain derived from the original graph. The existence of this example suggests that many more approaches discussed in this survey may be successfully applied to problems of knowledge discovery and data mining.

\section{Conclusion}
In this survey, we provided an overview on information-theoretic methods for coarse graining and model reduction for stationary, discrete-time Markov chains. We have focused on methods that quantify the operational goals of the respective problem (i.e., that the coarse graining is informative about a quantity of interest and that the reduced model is similar to the original model, respectively) using information-theoretic quantities. We have furthermore discussed the concept of lumpability, where the problems of coarse graining and model reduction intersect.

This survey shall act as a stepping stone from which the reader can explore the related literature on reduction of continuous-time Markov chains or general dynamical systems. Furthermore, since many settings in unsupervised machine learning can be parameterized as random walks, we believe that several of the approaches and methods presented in this survey can be successfully applied in knowledge discovery problems of various disciplines.

\appendices
\section{Proof of Theorem~\ref{thm:hmmaggregation}}\label{proof:hmmaggregation}
We need the following lemma.
\begin{lem}[{\cite[Thm.~13.8.1]{Cover_Information}}]\label{lem:hmm:approx}
\begin{multline}
  \pmf{Y}(y)=\sum_{x\in\set{X}} \pmf{X,Y}(x,y)\\ = \arg\min_{r_Y(y)} \expec{\kld{\pmf{Y|X}(\cdot|X)}{r_Y(\cdot)}}
\end{multline} 
\end{lem}

We now get
\begin{align}
 &\kld{\stoch{Y}_X,\stoch{Y}_O}{\tilde{\stoch{X}}',\tilde{\stoch{O}}'} \notag\\
 &= \lim_{T\to\infty} \frac{1}{T} \kld{\pmf{Y_{X,1}^T,Y_{O,1}^T}}{\pmf{\tilde X_1^{'T},\tilde O_1^{'T}}}\\
 &= \lim_{T\to\infty} \frac{1}{T} 
  \kld{\pmf{Y_{X,1}^T}}{\pmf{\tilde X_1^{'T}}} \notag\\
  &\quad+ 
  \expec{
    \kld{\pmf{Y_{O,1}^T|Y_{X,1}^T}(\cdot|Y_{X,1}^T)}  {\pmf{\tilde O_1^{'T}|\tilde X_1^{'T}}(\cdot|\tilde X_1^{'T})}
    } \label{eq:hmm:decomposition}.
\end{align}
The first term obviously only depends on $\tilde\tpm$. We write
\begin{align}
 &\min_{\tilde\tpm,W} \kld{\pmf{Y_{X,1}^T}}{\pmf{\tilde X_1^{'T}}}\notag\\
 &= \min_{\tilde\tpm} \sum_{t=1}^T \expec{\kld{\pmf{Y_{X,t}|Y_{X,1}^{t-1}}(\cdot|Y_{X,1}^{t-1})}{\tilde\tpm_{\tilde X_{t-1},\cdot}}}\\
 &\ge \sum_{t=1}^T \min_{\tilde\tpm_t} \expec{\kld{\pmf{Y_{X,t}|Y_{X,1}^{t-1}}(\cdot|Y_{X,1}^{t-1})}{\tilde\tpm_{\tilde X_{t-1},\cdot}}}
\end{align}
where the inequality follows by allowing the transition matrix $\tilde\tpm$ to depend on the time $t$. We now split the expectation over $Y_{X,1}^{t-1}$ into an outer expectation over $Y_{X,t-1}$ and an inner expectation over $Y_{X,1}^{t-2}|Y_{X,t-1}$. By Lemma~\ref{lem:hmm:approx}, the inner expectation is maximized, for every $Y_{X,t-1}$, by setting
\begin{multline}
 \tilde\tpm_{\ell,m}
 = \expec{\pmf{Y_{X,t}|Y_{X,1}^{t-2},Y_{X,t-1}}(m|Y_{X,1}^{t-2},\ell)}\\ = \pmf{Y_{X,t}|Y_{X,t-1}}(m|\ell).
\end{multline}
where the expectation is taken over $Y_{X,1}^{t-2}|Y_{X,t-1}=\ell$. Stationarity of $(\stoch{Y}_X,\stoch{Y}_O)$ ensures that the optimal $\tilde\tpm$ does not depend on the time $t$. This completes the first part of the proof.\endproof

For the second part, note that the second term in~\eqref{eq:hmm:decomposition} only depends on $W$:
\begin{align}
 &\min_{\tilde\tpm,W}  \expec{\kld{\pmf{Y_{O,1}^T|Y_{X,1}^T}(\cdot|Y_{X,1}^T)}  {\pmf{\tilde{O}_1^{T'}|\tilde{X}_1^{T'}}(\cdot|Y_{X,1}^T)}} \notag \\
 &= \min_W \expec{\kld{\pmf{Y_{O,1}^T|Y_{X,1}^T}(\cdot|Y_{X,1}^T)}  {\prod_{t=1}^T W_{Y_{X,t},\cdot}}} \\
 &\stackrel{(a)}{=} \min_W \sum_{t=1}^T \expec{
      \expec{
 \kld{\pmf{Y_{O,t}|Y_{X,1}^T,Y_{O,1}^{t-1}}(\cdot|Y_{X,1}^T,Y_{O,1}^{t-1})}{W_{Y_{X,t},\cdot}}
 }
 }\\
 &\stackrel{(b)}{\ge} \sum_{t=1}^T \min_{W_t} \expec{
      \expec{
 \kld{\pmf{Y_{O,t}|Y_{X,1}^T,Y_{O,1}^{t-1}}(\cdot|Y_{X,1}^T,Y_{O,1}^{t-1})}{W_{t,Y_{X,t},\cdot}}
 }
 }
\end{align}
where in $(a)$ the expectation is split into an inner expectation over $Y_{O,1}^{t-1}|Y_{X,1}^{T}$ and an outer expectation over $Y_{X,1}^{T}$, where in $(b)$ the outer expectation is w.r.t.\ $Y_{X,t}$ and the inner expectation is w.r.t.\ $Y_{X,1}^{t-1},Y_{X,t+1}^T,Y_{O,1}^{t-1}|Y_{X,t}$, and where the inequality follows by allowing the observation matrix $W$ to depend on the time $t$. By Lemma~\ref{lem:hmm:approx}, the inner expectation is maximized, for every $Y_{X,t}$, by setting
\begin{multline}
 W_{\ell,i}\\
 =  \expec{\pmf{Y_{O,t}|Y_{X,1}^{t-1},Y_{X,t},Y_{X,t+1}^T,Y_{O,1}^{t-1}}(i|Y_{X,1}^{t-1},\ell,Y_{X,t+1}^T,Y_{O,1}^{t-1})}\\
 = \pmf{Y_{O,t}|Y_{X,t}}(i|\ell).
\end{multline}
where the expectation is taken w.r.t.\ $Y_{X,1}^{t-1},Y_{X,t+1}^T,Y_{O,1}^{t-1}|Y_{X,t}$.
Stationarity of $(\stoch{Y}_X,\stoch{Y}_O)$ ensures that the optimal $W$ does not depend on the time $t$. This completes the proof.

\section*{Acknowledgments}
This work has been supported by the HiDALGO and BrAIN projects. The project HiDALGO (Grant  No.\ 824115) has been funded by the European Commission’s ICT activity of the H2020 Programme. The project BrAIN -- Brownfield Artificial Intelligence Network for Forging of High Quality Aerospace Components (FFG Grant No.\ 881039) is funded in the framework of the program ``TAKE OFF'', which is a research and technology program of the Austrian Federal Ministry of Transport, Innovation and Technology.

The Know-Center is funded within the Austrian COMET Program - Competence Centers for Excellent Technologies - under the auspices of the Austrian Federal Ministry of Climate Action, Environment, Energy, Mobility, Innovation and Technology, the Austrian Federal Ministry of Digital and Economic Affairs, and by the State of Styria. COMET is managed by the Austrian Research Promotion Agency FFG.

\bibliography{IEEEabrv,../../informationbottleneck/LiteratureSurvey/network_pruning,../../generalizedmarkovaggregation/references,../../../CV/myOwn}

\bibliographystyle{IEEEtran}

\end{document}